\documentclass[epj]{svjour}
\usepackage{latexsym}
\usepackage{feynmf}
\usepackage{amsmath,amssymb,amsfonts}
\begin{document}
\title{Renormalization of QED in an external field}
\author{Christian Brouder}
\institute{Laboratoire de Min\'eralogie-Cristallographie, 
  CNRS UMR7590, UPMC/UDD/IPGP, Case 115, 4 place Jussieu\\
  \small 75252 Paris {\sc cedex} 05, France\\
  \small \tt brouder@lmcp.jussieu.fr}
\date{\today}
\abstract{
The Schwinger equations of QED are rewritten
in three different ways as integral equations 
involving functional derivatives, which are
called weak field, strong field, and SCF quantum
electrodynamics. The perturbative solutions of
these equations are given in terms of appropriate
Feynman diagrams. The Green function that is
used as an electron propagator in each case is
discussed in detail. 
The general renormalization
rules for each of the three equations are
provided both in a non perturbative way
(Dyson relations) and for Feynman diagrams.
} 
\maketitle
\newcommand{\opphi}{\hat\phi}
\newcommand{\dd}{{\mathrm{d}}}
\newcommand{\ee}{{\mathrm{e}}}
\newcommand{\tr}{{\mathrm{tr}}}
\newcommand{\bfx}{{\mathbf{x}}}
\newcommand{\bfy}{{\mathbf{y}}}
\newcommand{\zbf}{{\mathbf{z}}}
\newcommand{\Ebf}{{\mathbf{E}}}
\newcommand{\Bbf}{{\mathbf{B}}}
\newcommand{\Abf}{{\mathbf{A}}}
\newcommand{\Tcal}{{\mathcal{T}}}
\newcommand{\Ccal}{{\mathcal{C}}}
\newcommand{\claphi}{\varphi_c}
\newcommand{\barA}{{\bar A}}
\newcommand{\bara}{{\bar a}}
\newcommand{\barj}{\bar {\text{\it\j}}}
\newcommand{\barJ}{\bar J} 
\newcommand{\barR}{\bar R}
\newcommand{\barS}{\bar S}
\newcommand{\barD}{\bar D}
\newcommand{\barPi}{\bar \Pi}
\newcommand{\psibf}{{\boldsymbol{\psi}}}
\newcommand{\Jbf}{{\mathbf{J}}}

\begin{fmffile}{dez4}
\setlength{\unitlength}{1mm}
\newcommand{\setval}{\fmfset{wiggly_len}{1.5mm}
\fmfset{arrow_len}{2.5mm}
\fmfset{arrow_ang}{13}\fmfset{dash_len}{1.5mm}\fmfpen{0.25mm}
\fmfset{dot_size}{0.8thick}}
\newcommand{\setgras}{\fmfset{wiggly_len}{1.5mm}
\fmfset{arrow_len}{2.5mm}
\fmfset{arrow_ang}{13}\fmfset{dash_len}{1.5mm}\fmfpen{0.5mm}
\fmfset{dot_size}{0.8thick}}
\newcommand{\scs}{\scriptstyle}

\section{Introduction}

This paper is a step towards the calculation of photon
and electron spectroscopies of matter based on 
quantum electrodynamics.

Starting from quantum electrodynamics, which is a 
most accurate and successful theory, garanties that the
basis of the calculation is sound.  Although such
an approach may look too true to be beautiful, it seems
adequate because of experimental and theoretical reasons.
From the experimental point of view, spectroscopy has made
huge advances and is now able to measure tiny effects.
From the theoretical point of view, relativistic methods
have proved very powerful, even for problems that do no
look obviously relativistic. For instance,
relativistic density functional theory 
\cite{Engel},\cite{Engel2} does not
meet the nonuniqueness problem of spin-density
functional theory \cite{Capelle}.
Moreover, relativistic quantum field theory
is a safe framework to go beyond the
present LDA methods, especially when dealing with
excitations.

There are many presentations of QED. We first have to
choose the version which is most convenient for 
solid-state physics. 
The main three formalisms used in textbooks 
are the quantum field, path integrals and Schr\"odinger
representations \cite{Hatfield}. The quantum field and
Schr\"odinger approaches manipulate operator-valued distributions
\cite{Wightman}
and the path integral approach uses ill-defined measures.
Therefore, we prefer to use Schwinger's idea Ref.\cite{Schwinger}
of working directly
with Green functions which are given as solutions of 
equations involving functional derivatives.

This approach has the mathematical advantage of manipulating
standard distributions, and the physical advantage of
using only measurable quantities, or the more measurable ones.

The standard approach to QED comes clearly from 
particle physics, where the S matrix is most useful, and
where in and out states of the scattering experiments are
well defined. In solid state physics, measurements are
usually based on a different principle. The spectroscopist
shines on the sample a beam of electrons or photons coming 
from a classical source.
By classical we mean that the source is not influenced
by the system being measured. After its interaction with
the sample, another beam of electrons or photons is measured.
Many spectroscopies can be described within this framework:
photoemission, electron scattering, photon scattering,
inelastic scattering, BIS, LEED, RHEED, x-ray absorption,
UV/visible spectroscopies, etc.
We try to stick as much as possible to this experimental
point of view.

We describe the photon field with
the electromagnetic potential (which is not
directly measurable but can be calculated in a given
gauge from the measurable
electric and magnetic fields) and the photon Green function
(from which the photon energy density can be computed).
We describe the electron field with
the one-particle Green function. This is also
not directly measurable, but many diagonal 
matrix elements of the Green function can be measured
(e.g. the electronic charge and current densities).

For applications to the spectroscopy of matter,
we need to formulate QED with an external field. This
external field is made of the potential due to the
nuclei, to magnetic or electric fields applied to
the sample and to external light sources.
After this introduction, we start with the definition
of the notation used for the Green functions and the
QED Lagrangian. Then, we derive the Schwinger equations
for the electromagnetic potential and the electron
propagator. These equations are solved iteratively
using three different methods (weak field, strong field,
self-consistent field). The boundary conditions,
such as the number of electrons in the system,
are determined by the unperturbed Green function
which is discussed in detail. The proper definition
of the current is established through the use of
a bilocal operator. Then we describe the renormalization
of QED in an external field. The non perturbative Dyson
relations are given to express the relation between bare
and renormalized Green functions. The
renormalization rules for Feynman diagrams are
discussed.

The following assumptions 
are made in the present paper: the external field
is weak enough not to create charges (this is true
for all stable atomic nuclei \cite{Greiner}) and the external
field is zero at infinity (we exclude constant
electric and magnetic field, for which specialized
monographs are available \cite{Grib},\cite{Dittrich}).
Moreover, we do not consider IR divergences.

\section{Notation}
In this section, we specify the notation that is used
in the paper.
The charge of the electron is $e=-|e|$.
The pseudo-metric tensor $g_{\lambda\mu}$ is
\begin{eqnarray*}
g=\left( \begin{array}{cccc} 1 & 0 & 0 & 0\\
                            0 &-1 & 0 & 0\\
                            0 & 0 &-1 & 0\\
                            0 & 0 & 0 &-1
                          \end{array} \right).
\end{eqnarray*}
We choose the standard gamma matrices
(\cite{Itzykson} p.693)
\begin{eqnarray*}
\gamma^0 &=&\left( \begin{array}{cc} I & 0 \\
                                  0 &-I \end{array} \right)
\quad
\gamma^j =\left( \begin{array}{cc} 0 & \sigma^j \\
                                 -\sigma^j & 0 \end{array}
\right).
\end{eqnarray*}
The Pauli matrices $\sigma^j$ are
defined by
\begin{eqnarray*}
\sigma^1=\left( \begin{array}{cc} 0 & 1 \\
                                 1 & 0 \end{array} \right)
\quad
\sigma^2=\left( \begin{array}{cc} 0 &-i \\
                                 i & 0 \end{array} \right)
\quad
\sigma^3=\left( \begin{array}{cc} 1 & 0 \\
                                 0 &-1 \end{array} \right).
\end{eqnarray*}
The charge conjugation and mass reversal matrices
are, respectively
\begin{eqnarray*}
C=\left( \begin{array}{cc} 0 & -i\sigma^2 \\
                                 -i\sigma^2 & 0 \end{array} \right),
\quad\mathrm{and}\quad
\gamma_5 = \left( \begin{array}{cc} 0 & I \\
                                 I & 0 \end{array} \right).
\end{eqnarray*}
They satisfy the following identities
\begin{eqnarray*}
C &=& -C^T = -C^\dagger = -C^{-1},\quad
C\gamma_\mu C^{-1} = -\gamma_\mu^T,\\
\gamma_5^{-1} &=& \gamma_5=\gamma^5,\quad \gamma_5\gamma^\mu\gamma_5=-\gamma^\mu.
\end{eqnarray*}

\subsection{Free propagators}

The free photon Green function $D^0_{\mu\nu}(x,y)$ is defined
by its Fourier transform
\begin{eqnarray*}
D^0_{\mu\nu}(x,y) &=& \int \frac{\dd q}{(2\pi)^4} \exp[-iq.(x-y)]
   D^0_{\mu\nu}(q), 
\end{eqnarray*}
where 
\begin{eqnarray*}
D^0_{\mu\nu}(q) &=& -\frac{1}{\epsilon_0  (q^2+i\epsilon)}
\big( g_{\mu\nu} -(1-1/\xi) q_\mu q_\nu/q^2 \big).
\end{eqnarray*}
It is a solution of the equation
\begin{eqnarray*}
\epsilon_0 \big( \Box g^{\mu\nu} - (1-\xi)\partial^\mu\partial^\nu) 
D^0_{\nu\lambda}(x,y) &=& \delta^\mu_{\,\,\lambda}\delta(x-y).
\end{eqnarray*}
In the last equation the D'Alembertian $\Box$ and the
derivatives $\partial^\mu\partial^\nu$ act on
variable $x$.

The free electron propagator is
\begin{eqnarray*}
S^0(x,y) &=& \int \frac{\dd q}{(2\pi)^4}
  \ee^{-iq.(x-y)} S^0(q),
\end{eqnarray*}
where
\begin{eqnarray*}
S^0(q)={(\hbar c\gamma\cdot q-mc^2+i\epsilon)}^{-1}.
\end{eqnarray*}
It is a solution of the equation
\begin{eqnarray*}
(i\hbar c \gamma\cdot\partial -mc^2) S^0(x,y) &=& \delta(x-y).
\end{eqnarray*}

Analytic expressions for these (and other) propagators are
given in \cite{Ma} and \cite{Scharf} section 2.3.
Notice that these Green functions are neither advanced
nor retarded. Their physical meaning is discussed
in \cite{Fierz}, \cite{StueckelbergR} and \cite{Pauli} p. 57.

\subsection{The QED Lagrangian in SI units}

To compare our results with those of nonrelativistic
many-body theory, it will be useful to write the
QED Lagrangian in SI units \cite{Cohen}.

We have $x^0=ct$, we
define the 4-current $J^\alpha = (\rho,\mathbf{J}/c)$,
and the 4-potential $A^\alpha = (V,c\mathbf{A})$.

The field-strength tensor is 
(\cite{Itzykson} p.8)
\begin{eqnarray*}
F^{\alpha\beta} &=& \partial^\alpha A^\beta -\partial^\beta A^\alpha
 = \left( \begin{array}{cccc}0 & -E^x & -E^y & -E^z\\
                               E^x & 0 & -cB^z & cB^y\\
                               E^y & cB^z & 0 & -cB^x\\
                               E^z & -cB^y & cB^x & 0\end{array} \right).
\end{eqnarray*}
The Maxwell equations are (\cite{Itzykson} p.9)
\begin{eqnarray*}
\partial_\alpha F^{\alpha\beta} &=& \frac{1}{\epsilon_0}
J^\beta,\\
\partial_\alpha \epsilon ^{\alpha\beta\gamma\delta} 
F_{\gamma\delta} &=&
0.
\end{eqnarray*}
The photon Lagrangian is (\cite{Itzykson} p.12)
\begin{eqnarray*}
{\cal {L}}_\gamma = -\frac{\epsilon_0}{4} F_{\alpha\beta}F^{\alpha\beta}
=\frac{\epsilon_0}{2} (|\Ebf|^2 - c^2|\Bbf|^2).
\end{eqnarray*}
The electron Lagrangian is
\begin{eqnarray*}
{\cal {L}}_e = i\hbar c \bar\psi \gamma\cdot\partial \psi
 -mc^2\bar\psi \psi,
\end{eqnarray*}
where $\gamma\cdot\partial=\gamma^\mu\partial_\mu$.
The interaction Lagrangian is
\begin{eqnarray*}
{\cal {L}}_I =  - J\cdot A =
-e \bar\psi \gamma\cdot A \psi.
\end{eqnarray*}
The current is
\begin{eqnarray*}
J_\mu(x) = e \bar\psi(x) \gamma_\mu \psi(x).
\end{eqnarray*}
The total Lagrangian in SI units, including the gauge term is
\begin{eqnarray*}
{\cal {L}} &=& -\frac{\epsilon_0}{4} F_{\alpha\beta}F^{\alpha\beta}
+\bar\psi(i\hbar c \gamma\cdot\partial-e\gamma\cdot A
-mc^2)\psi\\ &&-\frac{\epsilon_0\xi}{2}(\partial\cdot A)^2.
\end{eqnarray*}

\subsection{Dimensions}

The following table gives the SI units of 
the quantities used in the paper, in a
space-time with dimensions $d$ ($d=4$).

\begin{table}
\begin{tabular}{|l|l||l|l|}
$\epsilon_0$ & $L^{-3}M^{-1}T^2C^2$ &
$A^\alpha$ & $L^{4-d/2}MT^{-2}C^{-1}$ \\
$J^\alpha$ & $L^{1-d}C$ &
${\cal{L}}$ & $L^{3-d}MT^{-2}$ \\
$\psi$ & $L^{(1-d)/2}$ &
$e$ & $L^{d/2-2}C$ \\
$D$ & $L^{3-d/2}MT^{-2}C^{-2}$ &
$S$ & $M^{-1}L^{-d-2}T^2$ \\
$\Pi$ & $L^{-3-3d/2}M^{-1}T^2C^2$ &
$\Sigma$ & $ML^{2-d}T^{-2}$ \\
$\delta\Sigma/\delta a$ & $L^{-2-3d/2}C$ &
$\xi$ & 1 \\
$\hbar c$ & $ML^3T^{-2}$ &
$\eta$ & $ML^{(5-d)/2}T^{-2}$ \\
$I$  & $ML^3T^{-2}$ &
 & 
\end{tabular}
\end{table}

\section{Derivation of the functional equations}
The Schwinger equations were presented in Ref.\cite{Schwinger},
and various derivations of them are available
\cite{Anderson}, \cite{Bogoliubov} (p. 416-32) and
\cite{Itzykson} (p. 475-81). Our derivation
follows Ref.\cite{Itzykson}.

We define a generating function
$Z=Z(j,\eta,\bar\eta)$, where the photon source $j_\mu(x)$
and the anticommuting electron sources $\eta(x)$ and
$\bar\eta(x)$ enter the total action as \cite{Itzykson}
\begin{eqnarray*}
\sigma=I - \int \dd x\, j_\mu(x) A^\mu(x)
 +\int \dd x\, \bar\eta(x)\psi(x)+\bar\psi(x)\eta(x),
\end{eqnarray*}
where $I=\int \dd x {\cal {L}}(x)$.  
The minus sign before $j_\mu(x)$ in the definition of
$\sigma$ was chosen so that $j_\mu(x)$ is
a standard electromagnetic current.
It is the opposite of the convention used 
in Ref.\cite{Itzykson}. 

The generating
function can be written, up to a normalization factor, as a path integral
(\cite{Itzykson} p.476)
\begin{eqnarray}
Z(j,\bar\eta,\eta) &=& \int {\cal D}(A,\psi,\bar\psi)
\ee^{i\sigma/\hbar c},
\label{defZ}
\end{eqnarray}
or as the mean value of an operator
(\cite{Itzykson} p.210, 261)
\begin{eqnarray*}
Z(j,\bar\eta,\eta) &=& 
\langle T \exp\big( i(\sigma-I)/\hbar c\big)\rangle.
\end{eqnarray*}

It is also possible to write $Z(j,\bar\eta,\eta)$
in terms of a $Z_0(j,\bar\eta,\eta)$ without
interaction \cite{Itzykson} p. 445
and \cite{Ticciati} p.246.
Anyway, we do not calculate $Z$,
we only need it to calculate
mean values of Heisenberg operators.
For instance
\begin{eqnarray*}
Z \langle A_\mu(x) \rangle &=&
\int {\cal D}(A,\psi,\bar\psi) A_\mu(x)
\ee^{i\sigma/\hbar c}.
\end{eqnarray*}
From the definition of $\sigma$,
$A_\mu(x)$ can be written as a functional derivative
with respect to $j^\mu(x)$:
\begin{eqnarray}
Z \langle A_\mu(x) \rangle  &=&
-\int {\cal D}(A,\psi,\bar\psi)
\frac{\delta\sigma}{\delta j^\mu(x)}
\ee^{i\sigma/\hbar c}\nonumber\\
&=&
i\hbar c \frac{\delta Z}{\delta j^\mu(x)}.
\label{exemple}
\end{eqnarray}

Similarly, we shall use
\begin{eqnarray*}
Z \langle T \psi_\alpha(x) \bar\psi_\beta(y)\rangle  &=&
(\hbar c)^2
\frac{\delta^2 Z}
{\delta \eta^\beta(y)\delta\bar\eta^\alpha(x)},
\end{eqnarray*}
and
\begin{eqnarray*}
Z \langle T A_\lambda(x) A_\mu(y) \rangle &=&
-(\hbar c)^2
\frac{\delta^2 Z}{\delta j^\mu(y) \delta j^\lambda(x)}.
\end{eqnarray*}

Finally, the following property will be essential
\begin{eqnarray*}
\langle \psi(x) \rangle &=&
\langle \bar\psi(x) \rangle = 0.
\end{eqnarray*}
These equations are derived from the fact that the
ground state of the system is an eigenstate of the
charge. 
If $Q$ is the charge operator,
then $[Q,\psi(x)]=-\psi(x)$
(\cite{Itzykson} p.147), thus
\begin{eqnarray*}
\langle \psi(x) \rangle &=&
\langle G|\psi(x)|G \rangle =
-\langle G|Q \psi(x)|G \rangle
+\langle G|\psi(x) Q|G \rangle.
\end{eqnarray*}
The ground state is an eigenstate of $Q$
with eigenvalue $N$, thus
\begin{eqnarray*}
\langle G|\psi(x)|G \rangle &=&
-N\langle G|\psi(x)|G \rangle
+N\langle G|\psi(x)|G \rangle =0.
\end{eqnarray*}                                                                                 
More physically, no anticommuting operator
is measurable, only products of an even number
of anticommuting operators can be observed.
It may be stressed that the mean values of 
$\psi(x)$ and $\bar\psi(x)$ are zero
only when the
(unphysical) electron sources
$\eta(x)$ and $\bar\eta(x)$
are set to zero.

\subsection{The photon equation}

To derive the Schwinger equations 
we use the fact that the derivative of an integral is zero
(assuming that the integrand is zero at infinity):
\begin{eqnarray*}
\int {\cal D}(A,\psi,\bar\psi) 
\frac{\delta\ee^{i\sigma/\hbar c}}{\delta A^\mu(x)}=0.
\end{eqnarray*}
Thus
\begin{eqnarray}
\int {\cal D}(A,\psi,\bar\psi) \Big(\frac{\delta I}{\delta A^\mu(x)}
  -j_\mu(x)\Big) 
\ee^{i\sigma/\hbar c} =0.
\label{variA}
\end{eqnarray}
A direct calculation leads to
\begin{eqnarray}
\frac{\delta I}{\delta A^\mu(x)} &=&
\epsilon_0 \big( \Box g_{\mu\nu} - 
(1-\xi)\partial_\mu\partial_\nu\big) A^\nu(x)
\nonumber\\&&
-e \bar\psi(x)\gamma_\mu\psi(x).
\label{dIdA}
\end{eqnarray}

According to the method of generating
functions, a factor $A^\nu(x)$ in the
integral can be replaced 
by a functional derivative
of $Z$ with respect to $j_\nu(x)$ (up to a factor $i\hbar c$).
This is what we did for Eq.(\ref{exemple}).
In the case of anticommuting
variables such as $\psi(x)$ we must be a little
more careful.
Functional derivative with respect to an anticommuting source
is very similar to that with respect to a function.
The difference can be summarized in the identity
\begin{eqnarray*}
\frac{\delta}{\delta\eta} \big( AB) &=&
\frac{\delta A}{\delta\eta} B + (-1)^{|A|} A\frac{\delta B}{\delta\eta},
\end{eqnarray*}
where $A$ is a product of $|A|$ anticommuting variables.
For instance, $|A_\mu|=0$, $|\psi|=1$, $|\bar\eta|=1$, 
$|\bar\psi\gamma^\mu\psi|=2$, etc.

Thus, each factor $\psi(x)$ (resp. $\bar\psi(x)$)
 in the path integral is replaced by
a functional derivative of $Z$ with respect
to $\bar\eta(x)$ (resp. $\eta(x)$). Paying attention
to the signs and the factors $\hbar c/i$ we 
can rewrite Eq.(\ref{variA}) as
\begin{eqnarray}
\frac{\delta I}{\delta A^\mu(x)}
\Big(\frac{-\hbar c\delta}{i\delta j},
\frac{\hbar c\delta}{i\delta \bar\eta},
\frac{-\hbar c\delta}{i\delta \eta}\Big)Z = Z j_\mu(x).
\label{deltaIdeltaj}
\end{eqnarray}

Therefore, Eqs.(\ref{deltaIdeltaj}) and 
(\ref{dIdA}) in Eq.(\ref{variA}) yield
\begin{eqnarray*}
\epsilon_0 \big( \Box g_{\mu\nu} - (1-\xi)\partial_\mu\partial_\nu) 
\frac{-\hbar c\delta Z}{i\delta j_\nu(x)}\\
-e \frac{-\hbar c\delta}{i\delta \eta^s(x)}\gamma_\mu^{ss'}
\frac{\hbar c\delta Z}{i\delta \bar\eta^{s'}(x)} &=& Z j_\mu(x).
\end{eqnarray*}

The value of the vector potential and the electron wavefunctions are
obtained by \cite{Itzykson}
\begin{eqnarray}
A_\mu(x) &=& -\frac{\hbar c}{i} \frac{1}{Z} 
\frac{\delta Z}{\delta j^\mu(x)},
\label{defA}\\
\psi_s(x) &=& \frac{\hbar c}{i} \frac{1}{Z} 
\frac{\delta Z}{\delta \bar\eta^s(x)},
\label{defpsi}\\
\bar\psi_s(x) &=& -\frac{\hbar c}{i} \frac{1}{Z} 
\frac{\delta Z}{\delta \eta^s(x)}.
\label{defpsibar}
\end{eqnarray}
To simplify the notation, we have written
$A_\mu(x)$, $\psi_s(x)$ and $\bar\psi_s(x)$
for 
$\langle A_\mu(x) \rangle$,
$\langle \psi_s(x) \rangle$
and
$\langle \bar\psi_s(x) \rangle$.

Thus, we obtain
\begin{eqnarray*}
Z \epsilon_0 \big( \Box g_{\mu\nu} - (1-\xi)\partial_\mu\partial_\nu) 
A^\nu(x) \\
-e i\hbar c \frac{\delta}{\delta \eta^s(x)}\gamma_\mu^{ss'}
Z \psi_{s'}(x) &=& Z j_\mu(x).
\end{eqnarray*}
We define now the electron Green function by 
\begin{eqnarray}
S_{ss'}(x,y) &=& 
i\hbar c \frac{\delta^2 \log Z}{\delta \eta^{s'}(y)\delta \bar\eta^s(x)}
\nonumber\\ &=& -\frac{\delta \psi_s(x)}{\delta \eta^{s'}(y)}=
-\frac{\delta \bar\psi_{s'}(y)}{\delta \bar\eta^{s}(x)}.
\label{defSfunc}
\end{eqnarray}
Thus, the equation becomes

\begin{eqnarray*}
\epsilon_0 \big( \Box g_{\mu\nu} - (1-\xi)\partial_\mu\partial_\nu) 
A^\nu(x) \\
-e \bar\psi(x)\gamma_\mu\psi(x)
+ie\hbar c \tr[\gamma_\mu S(x,x)]
&=& j_\mu(x).
\end{eqnarray*}

When the external electron sources $\eta$ and
$\bar\eta$ are set to zero, we showed that $\psi(x)=\bar\psi(x)=0$, 
and we obtain our first basic equation
\begin{eqnarray}
\epsilon_0 \big( \Box g_{\mu\nu} - (1-\xi)\partial_\mu\partial_\nu) 
A^\nu(x) \nonumber\\
+ie\hbar c \tr[\gamma_\mu S(x,x)]
&=& j_\mu(x). \label{eqA}
\end{eqnarray}

It is a bit clumsy to use an electron source $\eta(x)$ just to
conclude that no such source exists which leads to the cancelation
of $\psi(x)$ and $\bar\psi(x)$. The way out of this difficulty
is to use Rochev's bilocal source $\eta(x,y)$ \cite{Rochev}.
\begin{eqnarray*}
\sigma=I - \int \dd x j_\mu(x) A^\mu(x)
 +\int \dd x \dd y \bar\psi(x) \eta(x,y) \psi(y),
\end{eqnarray*}
where $\eta(x,y)$ is now a physically reasonable source
of electron-positron pairs. Such a source would lead 
immediately to Eq.(\ref{eqA}). We used the more standard
electron sources to follow the textbook derivations \cite{Itzykson}.

Equation (\ref{eqA}) means physically that an induced
current $-ie\hbar c \tr[\gamma_\mu S(x,x)]$
must be added to the external current $j_\mu(x)$
as a source of electromagnetic potential.

\subsection{The electron equation}

The second equation is obtained by varying 
$\bar\psi(x)$.
Following \cite{Itzykson} p.478 we obtain

\begin{eqnarray}
\frac{\delta I}{\delta \bar\psi(x)}\Big(\frac{-\hbar c\delta}{i\delta j},
\frac{\hbar c\delta}{i\delta \bar\eta},
\frac{-\hbar c\delta}{i\delta \eta}\Big)Z = -Z\eta(x).
\label{deltaIdeltapsi}
\end{eqnarray}
The functional derivative of the action yields
\begin{eqnarray}
\frac{\delta I}{\delta \bar\psi(x)} &=&
(i\hbar c \gamma\cdot\partial -mc^2) \psi(x) -e\gamma\cdot A(x)\psi(x).
\label{funcderbarpsi}
\end{eqnarray}
Introducing this into Eq.(\ref{deltaIdeltapsi}), we obtain
\begin{eqnarray*}
-Z\eta(x) &=& (i\hbar c \gamma\cdot\partial -mc^2) 
\frac{\hbar c\delta Z}{i\delta \bar\eta(x)} -e
\gamma_\mu \frac{-\hbar c\delta}{i\delta j_\mu(x)}
\frac{\hbar c\delta Z}{i\delta \bar\eta(x)}.
\end{eqnarray*}
From Eq.(\ref{defpsi}) we can write
\begin{eqnarray*}
\frac{\hbar c}{i} \frac{\delta Z}{\delta \bar\eta^s(x)}
&=& Z\psi_s(x).
\end{eqnarray*}
Hence
\begin{eqnarray*}
-\eta(x) &=& (i\hbar c \gamma\cdot\partial -mc^2) 
\psi(x) -ie\hbar c
\gamma_\mu \frac{\delta\psi(x)}{\delta j_\mu(x)}\\&&
-e\gamma\cdot A(x)\psi(x).
\end{eqnarray*}

Finally, we take the functional derivative of this equation
with respect to $\eta(y)$, we use Eq.(\ref{defSfunc}), the fact 
that $\psi(x)=0$ and we obtain our second basic equation
\begin{eqnarray}
\delta(x-y) &=&
(i\hbar c \gamma\cdot\partial -mc^2-e\gamma\cdot A(x))S(x,y) \nonumber\\&&
 -ie\hbar c\gamma_\mu
\frac{\delta S(x,y)}{\delta j_\mu(x)}.
\label{secondbasic}
\end{eqnarray}

Again, the same result can be obtained without electron
sources $\eta$, $\bar\eta$ by using the electron-positron
source $\eta(x,y)$ \cite{Rochev}.

The equation for the electromagnetic vector potential is
\begin{eqnarray}
A_\mu(x) &=& a_\mu(x)
 -ie\hbar c \int \dd s D^0_{\mu\nu}(x,s)
 \tr[\gamma^\nu S(s,s)],\quad \label{barepot}
\end{eqnarray}
where $a_\mu(x)$ is the external potential
created by the external current $j_\nu(x)$:
\begin{eqnarray*}
a_\mu(x)=\int \dd y D^0_{\mu\nu}(x,y) j^\nu(y).
\end{eqnarray*}

In terms of the external potential $a_\mu(x)$, the
functional equation (\ref{secondbasic}) becomes
\begin{eqnarray}
(i\hbar c \gamma\cdot\partial -mc^2-e\gamma\cdot A(x))S(x,y) &=& \delta(x-y)
\nonumber\\
&&\hspace*{-40mm} +ie\hbar c\gamma^\mu
\int \dd s \frac{\delta S(x,y)}{\delta a_{\lambda}(s)}
  D^0_{\lambda\mu}(s,x). 
  \label{bareelectrondif0}
\end{eqnarray}

On the other hand, we can take the functional derivative
of the action with respect to $\psi(x)$
\begin{eqnarray}
\frac{\delta I}{\delta \psi(x)} &=&
i\hbar c \partial^\mu\bar\psi(x)\gamma_\mu
+\bar\psi(x)(e\gamma\cdot A^\mu(x)+mc^2).
\end{eqnarray}
We can now repeat the calculation that was done starting from
Eq.(\ref{funcderbarpsi}).
This gives us another equation for the electron Green function
\begin{eqnarray}
\delta(x-y) &=&
-i\hbar c \partial^\mu_x S(y,x)\gamma_\mu
  -S(y,x)(e\gamma\cdot A(x)+mc^2)
 \nonumber\\&&
 -ie\hbar c
\frac{\delta S(y,x)}{\delta j_\mu(x)}\gamma_\mu.
\label{anothereq}
\end{eqnarray}

\subsection{Photon Green function}

For spectroscopic applications, it is useful to 
know the photon Green function
\begin{eqnarray*}
D_{\mu\nu}(x,y) &=& \frac{\delta A_\mu(x)}{\delta j^\nu(y)}.
\end{eqnarray*}
In physical terms, the photon Green function gives the linear
response of the electromagnetic potential to the variation
$\delta j^\nu(y)$ of the external source by
\begin{eqnarray*}
\delta A_\mu(x) &=& \int \dd y D_{\mu\nu}(x,y) \delta j^\nu(y).
\end{eqnarray*}

To obtain an equation for $D_{\mu\nu}(x,y)$, we solve Eq.(\ref{eqA})
for $A_\mu(x)$.
\begin{eqnarray*}
A_\mu(x) &=& \int \dd z D^0_{\mu\lambda}(x,z)
 \Big( j^\lambda(z)-ie\hbar c \tr[\gamma^\lambda S(z,z)]
 \Big).
\end{eqnarray*}
A functional derivative with respect to $j^\nu(y)$ gives us
the equation for $D_{\mu\nu}(x,y)$
\begin{eqnarray}
D_{\mu\nu}(x,y) &=& D^0_{\mu\nu}(x,y)
 -ie\hbar c  \int \dd z D^0_{\mu\lambda}(x,z)
 \tr[\gamma^\lambda \frac{\delta S(z,z)}{\delta j^\nu(y)}].
\nonumber\\&& \label{eqD1}
\end{eqnarray}

\section{Three integral equations}
In this section, we derive various integral equations
which correspond to the differential equations of the previous
section.

\subsection{Weak external potential}
When the external potential is weak, we can multiply
Eq.(\ref{bareelectrondif0}) by the free electron Green function $S^0(x,y)$.
This gives us the following coupled equations
\begin{eqnarray*}
S(x,y) &=& S^0(x,y)+ e\int \dd z S^0(x,z)\gamma\cdot A(z)S(z,y)\\&&
+ie\hbar c\int \dd z \dd z'  S^0(x,z) \gamma_\mu
\frac{\delta S(z,y)}{\delta a^{\nu}(z')}D^0_{\nu\mu}(z',z),\\
A_\mu(x) &=& a_\mu(x)
 -ie\hbar c \int \dd y D^0_{\mu\nu}(x,y)\tr[\gamma^\nu S(y,y)].
\end{eqnarray*}

The weak field approach to atomic physics was
reviewed recently by Eides and coll. \cite{Eides}. 

We denote the external current $j_\mu(x)$
by a star
$\parbox{4mm}{\begin{center}
\begin{fmfgraph}(2,2)
  \fmfforce{0.5w,0.5h}{vx}
\fmfv{decor.shape=pentagram,decor.filled=1,
      decor.size=2thick}{vx}
\end{fmfgraph}
\end{center}}$,
so the external potential $a_\mu(x)$
is denoted by the Feynman diagram
$\parbox{5mm}{\begin{center}
\begin{fmfgraph}(5,2)
\setval
\fmfforce{0w,0.5h}{v1}
\fmfforce{0.9w,0.5h}{v2}
\fmf{boson}{v1,v2}
\fmfdot{v1}
\fmfv{decor.shape=pentagram,decor.filled=1,
      decor.size=2thick}{v2}
\end{fmfgraph}
\end{center}}$\,.

\subsubsection{Feynman diagrams}
The coupled equations generate the
following series for the electron
Green function.
The propagator is oriented
from right to left and the electron loops
are oriented anticlockwise.

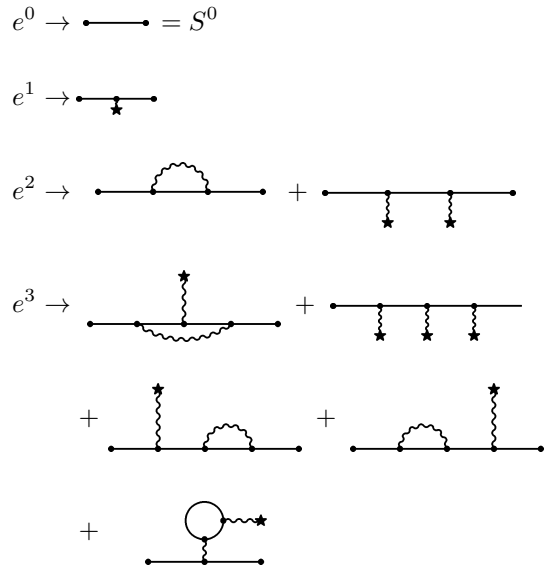
\begin{eqnarray*}
e^0 &\rightarrow&
  \parbox{10mm}{\begin{center}
  \begin{fmfgraph*}(8,3)
  \setval
  \fmfleft{v1}
  \fmfright{v2}
  \fmf{plain}{v2,v1}
  \fmfdot{v1,v2}
  \end{fmfgraph*}
  \end{center}} = S^0
\\
e^1 &\rightarrow&
  \parbox{10mm}{\begin{center}
  \begin{fmfgraph*}(10,3)
  \setval
  \fmfforce{0w,0.5h}{v1}
  \fmfforce{0.5w,0.0h}{vx}
  \fmfforce{0.5w,0.5h}{vy}
  \fmfforce{1w,0.5h}{v2}
  \fmf{plain}{v2,vy,v1}
  \fmf{boson}{vx,vy}
  \fmfdot{v1,v2,vy}
  \fmfv{decor.shape=pentagram,decor.filled=1,
      decor.size=2thick}{vx}
  \end{fmfgraph*}
  \end{center}}
\\
e^2 &\rightarrow&
 \parbox{27mm}{\begin{center}
  \begin{fmfgraph}(22,5)
  \setval
  \fmfforce{0w,0.5h}{v1}
  \fmfforce{1/3w,0.5h}{v2}
  \fmfforce{2/3w,0.5h}{v3}
  \fmfforce{1w,0.5h}{v4}
  \fmf{plain}{v4,v3,v2,v1}
  \fmf{boson,left}{v2,v3}
  \fmfdot{v1,v2,v3,v4}
  \end{fmfgraph}
  \end{center}}
+
\parbox{28mm}{\begin{center}
\begin{fmfgraph}(25,8)
\setval
  \fmfforce{0w,0.5h}{v1}
  \fmfforce{1/3w,0.5h}{v2}
  \fmfforce{2/3w,0.5h}{v3}
  \fmfforce{1/3w,0.0h}{v2x}
  \fmfforce{2/3w,0.0h}{v3x}
  \fmfforce{1w,0.5h}{v4}
  \fmf{plain}{v4,v3,v2,v1}
  \fmf{boson}{v2,v2x}
  \fmf{boson}{v3,v3x}
  \fmfdot{v1,v2,v3,v4}
  \fmfv{decor.shape=pentagram,decor.filled=1,
      decor.size=2thick}{v2x,v3x}
\end{fmfgraph}
\end{center}}
\\
e^3 &\rightarrow&
\parbox{28mm}{\begin{center}
\begin{fmfgraph}(25,8)
\setval
\fmfforce{0w,0.2h}{v1}
\fmfforce{0.25w,0.2h}{v2}
\fmfforce{0.5w,0.2h}{vx}
\fmfforce{0.5w,1h}{vxb}
\fmfforce{0.75w,0.2h}{v3}
\fmfforce{1w,00.2h}{v4}
\fmf{plain}{v4,v3,vx,v2,v1}
\fmf{boson,right=0.33}{v2,v3}
\fmf{boson}{vx,vxb}
\fmfdot{v1,v2,vx,v3,v4}
  \fmfv{decor.shape=pentagram,decor.filled=1,
      decor.size=2thick}{vxb}
\end{fmfgraph}
\end{center}}
+
\parbox{28mm}{\begin{center}
\begin{fmfgraph}(25,8)
\setval
  \fmfforce{0w,0.5h}{v1}
  \fmfforce{1/4w,0.5h}{v2}
  \fmfforce{2/4w,0.5h}{v3}
  \fmfforce{3/4w,0.5h}{v4}
  \fmfforce{1/4w,0.0h}{v2x}
  \fmfforce{2/4w,0.0h}{v3x}
  \fmfforce{3/4w,0.0h}{v4x}
  \fmfforce{1w,0.5h}{v5}
  \fmf{plain}{v5,v4,v3,v2,v1}
  \fmf{boson}{v2,v2x}
  \fmf{boson}{v3,v3x}
  \fmf{boson}{v4,v4x}
  \fmfdot{v1,v2,v3,v4}
  \fmfv{decor.shape=pentagram,decor.filled=1,
      decor.size=2thick}{v2x,v3x,v4x}
\end{fmfgraph}
\end{center}}
\\&&
+
\parbox{28mm}{\begin{center}
\begin{fmfgraph}(25,8)
\setval
\fmfforce{0.0w,0h}{v1}
\fmfforce{0.25w,0h}{v2}
\fmfforce{0.25w,1h}{v2x}
\fmfforce{0.5w,0h}{v3}
\fmfforce{0.75w,0h}{v4}
\fmfforce{1w,0h}{v5}
\fmf{plain}{v5,v4,v3,v2,v1}
\fmf{boson,left}{v3,v4}
\fmf{boson}{v2,v2x}
\fmfdot{v1,v2,v3,v4,v5}
\fmfv{decor.shape=pentagram,decor.filled=1,
      decor.size=2thick}{v2x}
\end{fmfgraph}
\end{center}}
+
\parbox{28mm}{\begin{center}
\begin{fmfgraph}(25,8)
\setval
\fmfforce{0.0w,0h}{v1}
\fmfforce{0.25w,0h}{v2}
\fmfforce{0.5w,0h}{v3}
\fmfforce{0.75w,0h}{v4}
\fmfforce{0.75w,1h}{v4x}
\fmfforce{1w,0h}{v5}
\fmf{plain}{v5,v4,v3,v2,v1}
\fmf{boson,left}{v2,v3}
\fmf{boson}{v4,v4x}
\fmfdot{v1,v2,v3,v4,v5}
\fmfv{decor.shape=pentagram,decor.filled=1,
      decor.size=2thick}{v4x}
\end{fmfgraph}
\end{center}}
\\&&
+
\parbox{28mm}{\begin{center}
\begin{fmfgraph}(25,8)
\setval
\fmfforce{0.2w,0h}{v2}
\fmfforce{0.4w,11/16h}{v2b}
\fmfforce{0.5w,0h}{vx}
\fmfforce{0.5w,6/16h}{vxb}
\fmfforce{0.8w,0h}{v3}
\fmfforce{0.6w,11/16h}{v3b}
\fmfforce{0.8w,11/16h}{vy}
\fmf{plain}{v3,vx,v2}
\fmf{boson}{vx,vxb}
\fmf{boson}{v3b,vy}
\fmf{plain,right}{v2b,v3b,v2b}
\fmfdot{v2,vx,vxb,v3,v3b}
\fmfv{decor.shape=pentagram,decor.filled=1,
      decor.size=2thick}{vy}
\end{fmfgraph}
\end{center}}
\end{eqnarray*}

\subsection{Strong external potential}
When the external potential is strong, we multiply 
Eq.(\ref{bareelectrondif0}) by
the Green function in the presence of $a_\mu(x)$:
\begin{eqnarray*}
S^a &=& \big(i\hbar c \gamma\cdot\partial -mc^2-e\gamma\cdot a(x)\big)^{-1}.
\end{eqnarray*}
Now the coupled equations become
\begin{eqnarray*}
S(x,y) &=& S^a(x,y)+ e\int \dd z S^a(x,z)\gamma\cdot A'(z)S(z,y)\\&&
+ie\hbar c\int \dd z \dd z'  S^a(x,z) \gamma_\mu
\frac{\delta S(z,y)}{\delta a^{\nu}(z')}D^0_{\nu\mu}(z',z),\\
A'_\mu(x) &=& 
 -ie\hbar c \int \dd s D^0_{\mu\nu}(x,y)\tr[\gamma^\nu S(y,y)].
\end{eqnarray*}
In this case, the total potential is $A_\mu(x)=a_\mu(x)+A'_\mu(x)$.

Strong field QED was reviewed recently in Ref. \cite{Mohr}. 
Strong field QED can easily accomodate bound states, but
the nuclear potential is not screened.

\subsubsection{Feynman diagrams \label{FeynStrong}}

For the electron Green functions, the propagator is oriented
from right to left and the electron loops
are oriented anticlockwise, except for the tadpoles, where
a loop diagram is half the sum of a clockwise loop and an 
anticlockwise loop. 

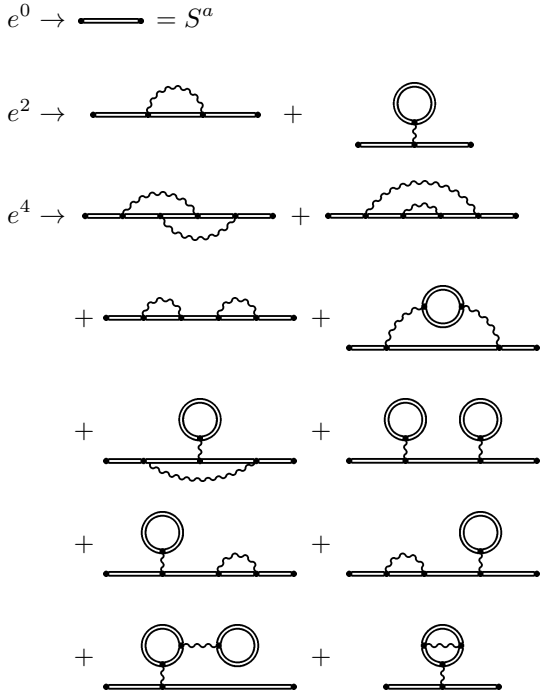
\begin{eqnarray*}
e^0 &\rightarrow&
  \parbox{10mm}{\begin{center}
  \begin{fmfgraph*}(8,3)
  \setval
  \fmfleft{v1}
  \fmfright{v2}
  \fmf{dbl_plain}{v2,v1}
  \fmfdot{v1,v2}
  \end{fmfgraph*}
  \end{center}} = S^a
\\
e^2 &\rightarrow&
 \parbox{27mm}{\begin{center}
  \begin{fmfgraph}(22,5)
  \setval
  \fmfforce{0w,0.5h}{v1}
  \fmfforce{1/3w,0.5h}{v2}
  \fmfforce{2/3w,0.5h}{v3}
  \fmfforce{1w,0.5h}{v4}
  \fmf{dbl_plain}{v4,v3,v2,v1}
  \fmf{boson,left}{v2,v3}
  \fmfdot{v1,v2,v3,v4}
  \end{fmfgraph}
  \end{center}}
+
\parbox{28mm}{\begin{center}
\begin{fmfgraph}(25,8)
\setval
\fmfforce{0.2w,0h}{v2}
\fmfforce{0.4w,11/16h}{v2b}
\fmfforce{0.5w,0h}{vx}
\fmfforce{0.5w,6/16h}{vxb}
\fmfforce{0.8w,0h}{v3}
\fmfforce{0.6w,11/16h}{v3b}
\fmf{dbl_plain}{v3,vx,v2}
\fmf{boson}{vx,vxb}
\fmf{dbl_plain,right}{v2b,v3b,v2b}
\fmfdot{v2,vx,vxb,v3}
\end{fmfgraph}
\end{center}}
\\
e^4 &\rightarrow&
\parbox{28mm}{\begin{center}
\begin{fmfgraph}(25,5)
\setval
\fmfforce{0w,0.5h}{v1}
\fmfforce{1/5w,0.5h}{v2}
\fmfforce{2/5w,0.5h}{v3}
\fmfforce{3/5w,0.5h}{v4}
\fmfforce{4/5w,0.5h}{v5}
\fmfforce{1w,0.5h}{v6}
\fmf{dbl_plain}{v6,v5,v4,v3,v2,v1}
\fmf{boson,left=0.6}{v2,v4}
\fmf{boson,right=0.6}{v3,v5}
\fmfdot{v1,v2,v3,v4,v5,v6}
\end{fmfgraph}
\end{center}}
+
\parbox{28mm}{\begin{center}
\begin{fmfgraph}(25,5)
\setval
\fmfforce{0w,0.5h}{v1}
\fmfforce{1/5w,0.5h}{v2}
\fmfforce{2/5w,0.5h}{v3}
\fmfforce{3/5w,0.5h}{v4}
\fmfforce{4/5w,0.5h}{v5}
\fmfforce{1w,0.5h}{v6}
\fmf{dbl_plain}{v6,v5,v4,v3,v2,v1}
\fmf{boson,left=0.6}{v2,v5}
\fmf{boson,left=0.6}{v3,v4}
\fmfdot{v1,v2,v3,v4,v5,v6}
\end{fmfgraph}
\end{center}}
\\&&
+
\parbox{28mm}{\begin{center}
\begin{fmfgraph}(25,5)
\setval
\fmfforce{0w,0.5h}{v1}
\fmfforce{1/5w,0.5h}{v2}
\fmfforce{2/5w,0.5h}{v3}
\fmfforce{3/5w,0.5h}{v4}
\fmfforce{4/5w,0.5h}{v5}
\fmfforce{1w,0.5h}{v6}
\fmf{dbl_plain}{v6,v5,v4,v3,v2,v1}
\fmf{boson,left}{v2,v3}
\fmf{boson,left}{v4,v5}
\fmfdot{v1,v2,v3,v4,v5,v6}
\end{fmfgraph}
\end{center}}
+
\parbox{28mm}{\begin{center}
\begin{fmfgraph}(25,8)
\setval
\fmfforce{0w,0h}{v1}
\fmfforce{0.2w,0h}{v2}
\fmfforce{0.4w,11/16h}{v2b}
\fmfforce{0.8w,0h}{v3}
\fmfforce{0.6w,11/16h}{v3b}
\fmfforce{1w,0h}{v4}
\fmf{dbl_plain}{v4,v3,v2,v1}
\fmf{boson,left=0.33}{v2,v2b}
\fmf{boson,right=0.33}{v3,v3b}
\fmf{dbl_plain,right}{v2b,v3b,v2b}
\fmfdot{v1,v2,v3,v3b,v2b,v4}
\end{fmfgraph}
\end{center}}
\\&&
+
\parbox{28mm}{\begin{center}
\begin{fmfgraph}(25,8)
\setval
\fmfforce{0w,0h}{v1}
\fmfforce{0.2w,0h}{v2}
\fmfforce{0.4w,11/16h}{v2b}
\fmfforce{0.5w,0h}{vx}
\fmfforce{0.5w,6/16h}{vxb}
\fmfforce{0.8w,0h}{v3}
\fmfforce{0.6w,11/16h}{v3b}
\fmfforce{1w,0h}{v4}
\fmf{dbl_plain}{v4,v3,v2,v1}
\fmf{boson,right=0.33}{v2,v3}
\fmf{boson}{vx,vxb}
\fmf{dbl_plain,right}{v2b,v3b,v2b}
\fmfdot{v1,v2,v3,vx,vxb,v4}
\end{fmfgraph}
\end{center}}
+
\parbox{28mm}{\begin{center}
\begin{fmfgraph}(25,8)
\setval
\fmfforce{0.0w,0h}{v2}
\fmfforce{0.2w,11/16h}{v2b}
\fmfforce{0.3w,0h}{vx}
\fmfforce{0.3w,6/16h}{vxb}
\fmfforce{0.4w,11/16h}{v3b}
\fmfforce{0.6w,11/16h}{x2b}
\fmfforce{0.7w,0h}{xx}
\fmfforce{0.7w,6/16h}{xxb}
\fmfforce{0.8w,11/16h}{x3b}
\fmfforce{1w,0h}{v4}
\fmf{dbl_plain}{v4,xx,vx,v2}
\fmf{boson}{vx,vxb}
\fmf{boson}{xx,xxb}
\fmf{dbl_plain,right}{v2b,v3b,v2b}
\fmf{dbl_plain,right}{x2b,x3b,x2b}
\fmfdot{v2,vx,vxb,xx,xxb,v4}
\end{fmfgraph}
\end{center}}
\\&&
+
\parbox{28mm}{\begin{center}
\begin{fmfgraph}(25,8)
\setval
\fmfforce{0.0w,0h}{v2}
\fmfforce{0.2w,11/16h}{v2b}
\fmfforce{0.3w,0h}{vx}
\fmfforce{0.3w,6/16h}{vxb}
\fmfforce{0.4w,11/16h}{v3b}
\fmfforce{0.6w,0h}{y1}
\fmfforce{0.8w,0h}{y2}
\fmfforce{1w,0h}{v4}
\fmf{dbl_plain}{v4,y2,y1,vx,v2}
\fmf{boson}{vx,vxb}
\fmf{boson,left}{y1,y2}
\fmf{dbl_plain,right}{v2b,v3b,v2b}
\fmfdot{v2,vx,vxb,y1,y2,v4}
\end{fmfgraph}
\end{center}}
+
\parbox{28mm}{\begin{center}
\begin{fmfgraph}(25,8)
\setval
\fmfforce{0.0w,0h}{v2}
\fmfforce{0.2w,0h}{y1}
\fmfforce{0.4w,0h}{y2}
\fmfforce{0.6w,11/16h}{x2b}
\fmfforce{0.7w,0h}{xx}
\fmfforce{0.7w,6/16h}{xxb}
\fmfforce{0.8w,11/16h}{x3b}
\fmfforce{1w,0h}{v4}
\fmf{dbl_plain}{v4,xx,y2,y1,v2}
\fmf{boson,left}{y1,y2}
\fmf{boson}{xx,xxb}
\fmf{dbl_plain,right}{x2b,x3b,x2b}
\fmfdot{v2,y1,y2,xx,xxb,v4}
\end{fmfgraph}
\end{center}}
\\&&+
\parbox{28mm}{\begin{center}
\begin{fmfgraph}(25,8)
\setval
\fmfforce{0.0w,0h}{v2}
\fmfforce{0.2w,11/16h}{v2b}
\fmfforce{0.3w,0h}{vx}
\fmfforce{0.3w,6/16h}{vxb}
\fmfforce{0.4w,11/16h}{v3b}
\fmfforce{0.6w,11/16h}{x2b}
\fmfforce{0.7w,0h}{xx}
\fmfforce{0.7w,6/16h}{xxb}
\fmfforce{0.8w,11/16h}{x3b}
\fmfforce{1w,0h}{v4}
\fmf{dbl_plain}{v4,vx,v2}
\fmf{boson}{vx,vxb}
\fmf{boson}{v3b,x2b}
\fmf{dbl_plain,right}{v2b,v3b,v2b}
\fmf{dbl_plain,right}{x2b,x3b,x2b}
\fmfdot{v2,vx,vxb,v3b,x2b,v4}
\end{fmfgraph}
\end{center}}
+
\parbox{28mm}{\begin{center}
\begin{fmfgraph}(25,8)
\setval
\fmfforce{0.2w,0h}{v2}
\fmfforce{0.4w,11/16h}{v2b}
\fmfforce{0.5w,0h}{vx}
\fmfforce{0.5w,6/16h}{vxb}
\fmfforce{0.8w,0h}{v3}
\fmfforce{0.6w,11/16h}{v3b}
\fmf{dbl_plain}{v3,vx,v2}
\fmf{boson}{vx,vxb}
\fmf{boson}{v2b,v3b}
\fmf{dbl_plain,right}{v2b,v3b,v2b}
\fmfdot{v2,vx,vxb,v3,v2b,v3b}
\end{fmfgraph}
\end{center}}
\end{eqnarray*}

For the potential $A^\mu(x)$ we have the following
Feynman diagrams

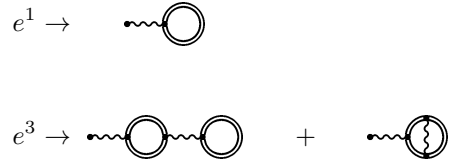
\begin{eqnarray*}
e^1 &\rightarrow&
\parbox{28mm}{\begin{center}
\begin{fmfgraph}(25,8)
\setval
\fmfforce{0.2w,0.5h}{v2}
\fmfforce{0.4w,8/16h}{v2b}
\fmfforce{0.6w,8/16h}{v3b}
\fmf{boson}{v2,v2b}
\fmf{dbl_plain,right}{v2b,v3b,v2b}
\fmfdot{v2,v2b}
\end{fmfgraph}
\end{center}}
\\
e^3 &\rightarrow&
\parbox{28mm}{\begin{center}
\begin{fmfgraph}(25,8)
\setval
\fmfforce{0.0w,0.5h}{v2}
\fmfforce{0.2w,8/16h}{v2b}
\fmfforce{0.4w,8/16h}{v3b}
\fmfforce{0.6w,8/16h}{x2b}
\fmfforce{0.8w,8/16h}{x3b}
\fmf{boson}{v2,v2b}
\fmf{boson}{v3b,x2b}
\fmf{dbl_plain,right}{v2b,v3b,v2b}
\fmf{dbl_plain,right}{x2b,x3b,x2b}
\fmfdot{v2,v2b,v3b,x2b}
\end{fmfgraph}
\end{center}}
+
\parbox{28mm}{\begin{center}
\begin{fmfgraph}(25,8)
\setval
\fmfforce{0.2w,0.5h}{v2}
\fmfforce{0.4w,8/16h}{v2b}
\fmfforce{0.5w,3/16h}{vxb}
\fmfforce{0.5w,13/16h}{vxc}
\fmfforce{0.6w,8/16h}{v3b}
\fmf{boson}{v2,v2b}
\fmf{boson}{vxb,vxc}
\fmf{dbl_plain,right}{v2b,v3b,v2b}
\fmfdot{v2,v2b,vxb,vxc}
\end{fmfgraph}
\end{center}}
\end{eqnarray*}

If $s_n$ (resp. $a_n$)
denotes the number of Feynman diagrams for
$S(x,y)$ (resp $A^\mu(x)$) at order $e^n$,
the coupled equations yield
\begin{eqnarray*}
s_0&=&1, \quad a_0 = 0,\\
a_n&=& s_{n-1},\\
s_n&=&\sum_{i=0}^{n-1} s_i a_{n-i-1} +(n-1)s_{n-2}.
\end{eqnarray*}
Thus, 
$s_{2n}=1,2,10,74,706,8162,110410,...$ and
$s_{2n+1}=0$.
If $s(x)$ is the generating function for the sequence
$s_n$, then
\begin{eqnarray*}
s(x) &=& \sum_{n=0}^\infty s_n x^n = \frac{f(x)}{1+x^2 f(x)},
\end{eqnarray*}
where 
\begin{eqnarray*}
f(x) &=& \sum_{n=0}^\infty (2n+1)!!\, x^{2n}.
\end{eqnarray*}
For similar results, see \cite{Itzykson} p.467.

\subsection{SCF external potential}
Finally, the most accurate results are obtained when the 
external field is taken as the complete vector potential $A_\mu(x)$,
which is determined by a self-consistent field procedure.
Therefore, the potential that will be used in the
initial Green function is $A_\mu(x)$ instead of
$a_\mu(x)$. This calculation can be found
in Refs.\cite{Bogoliubov} and \cite{Itzykson} p.479.

To eliminate the external current $j^\nu(y)$
from Eq.(\ref{secondbasic})
we need the functional relation
\begin{eqnarray*}
\frac{\delta S(x,z)}{\delta j^\nu(y)} &=&
\int \dd s \frac{\delta S(x,z)}{\delta A_\mu(s)}
\frac{\delta A_\mu(s)}{\delta j^\nu(y)}\\
&=&\int \dd s \frac{\delta S(x,z)}{\delta A_\mu(s)}D_{\mu\nu}(s,y).
\end{eqnarray*}

With this relation, the functional equation for the
photon Green function Eq.(\ref{eqD1}) becomes
\begin{eqnarray}
D_{\mu\nu}(x,y) &=& D^0_{\mu\nu}(x,y)
 -ie\hbar c  \int \dd z \dd s D^0_{\mu\lambda}(x,z)\nonumber\\&&\times
 \tr[\gamma^\lambda \frac{\delta S(z,z)}{\delta A_{\lambda'}(s)}]
  D_{\lambda'\nu}(s,y), \label{barephoton}
\end{eqnarray}
and the functional equation for the electron Green function 
Eq.(\ref{secondbasic}) becomes
\begin{eqnarray}
(i\hbar c \gamma\cdot\partial -mc^2-e\gamma\cdot A(x))S(x,y) &=& \delta(x-y)
\nonumber\\
&&\hspace*{-40mm} +ie\hbar c\gamma^\mu
\int \dd s \frac{\delta S(x,y)}{\delta A_{\lambda}(s)}
  D_{\lambda\mu}(s,x). \label{bareelectrondif}
\end{eqnarray}

Then, we multiply Eq.(\ref{bareelectrondif}) by
the Green function in the presence of $A_\mu(x)$:
\begin{eqnarray*}
S^A &=& \big(i\hbar c \gamma\cdot\partial -mc^2-e\gamma\cdot A(x)\big)^{-1},
\end{eqnarray*}
and the equation for the electron Green function becomes
\begin{eqnarray}
S(x,y) &=& S^A(x,y)\nonumber\\&&\hspace*{-9mm}
+ie\hbar c\int \dd z \dd z'  S^A(x,z) \gamma_\mu
\frac{\delta S(z,y)}{\delta A^{\nu}(z')}D^{\nu\mu}(z',z).
\label{SCFQED}
\end{eqnarray}
This equation, together with Eq.(\ref{barephoton}) and 
Eq.(\ref{barepot}) are a complete system of equations for the
determination of the bare Green functions and the potential.

As compared to the case of a strong external potential,
the SCF potential has the advantage of taking into account
the electrons in the system. Thus, the nuclear potentials
are screened by the electrons in a self-consistent way.
All the tadpoles of strong field QED have been resummed.

This formulation of QED is closer to the standard methods
of solid state physics or quantum chemistry. It was 
used in nuclear physics under the name Hartree
QED when the current is calculated by a single electron
loop \cite{Hamm}, see also \cite{Bielajew}.

\subsubsection{Electron Green function for self-consistent field}

For the electron Green functions, all electron loops
are oriented anticlockwise and the propagator is oriented
from right to left.

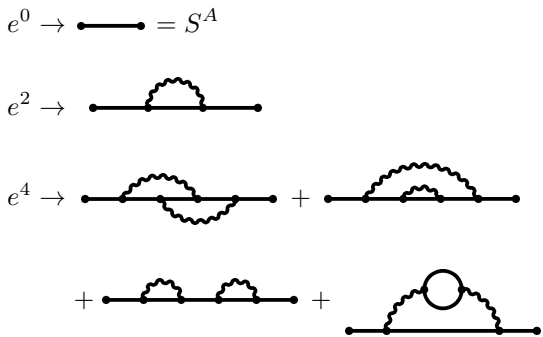
\begin{eqnarray*}
e^0 &\rightarrow&
  \parbox{10mm}{\begin{center}
  \begin{fmfgraph*}(8,3)
  \setgras
  \fmfleft{v1}
  \fmfright{v2}
  \fmf{plain}{v2,v1}
  \fmfdot{v1,v2}
  \end{fmfgraph*}
  \end{center}} = S^A
\\
e^2 &\rightarrow&
 \parbox{27mm}{\begin{center}
  \begin{fmfgraph}(22,5)
  \setgras
  \fmfforce{0w,0.5h}{v1}
  \fmfforce{1/3w,0.5h}{v2}
  \fmfforce{2/3w,0.5h}{v3}
  \fmfforce{1w,0.5h}{v4}
  \fmf{plain}{v4,v3,v2,v1}
  \fmf{boson,left}{v2,v3}
  \fmfdot{v1,v2,v3,v4}
  \end{fmfgraph}
  \end{center}}
\\
e^4 &\rightarrow&
\parbox{28mm}{\begin{center}
\begin{fmfgraph}(25,5)
\setgras
\fmfforce{0w,0.5h}{v1}
\fmfforce{1/5w,0.5h}{v2}
\fmfforce{2/5w,0.5h}{v3}
\fmfforce{3/5w,0.5h}{v4}
\fmfforce{4/5w,0.5h}{v5}
\fmfforce{1w,0.5h}{v6}
\fmf{plain}{v6,v5,v4,v3,v2,v1}
\fmf{boson,left=0.6}{v2,v4}
\fmf{boson,right=0.6}{v3,v5}
\fmfdot{v1,v2,v3,v4,v5,v6}
\end{fmfgraph}
\end{center}}
+
\parbox{28mm}{\begin{center}
\begin{fmfgraph}(25,5)
\setgras
\fmfforce{0w,0.5h}{v1}
\fmfforce{1/5w,0.5h}{v2}
\fmfforce{2/5w,0.5h}{v3}
\fmfforce{3/5w,0.5h}{v4}
\fmfforce{4/5w,0.5h}{v5}
\fmfforce{1w,0.5h}{v6}
\fmf{plain}{v6,v5,v4,v3,v2,v1}
\fmf{boson,left=0.6}{v2,v5}
\fmf{boson,left=0.6}{v3,v4}
\fmfdot{v1,v2,v3,v4,v5,v6}
\end{fmfgraph}
\end{center}}
\\&&
+
\parbox{28mm}{\begin{center}
\begin{fmfgraph}(25,5)
\setgras
\fmfforce{0w,0.5h}{v1}
\fmfforce{1/5w,0.5h}{v2}
\fmfforce{2/5w,0.5h}{v3}
\fmfforce{3/5w,0.5h}{v4}
\fmfforce{4/5w,0.5h}{v5}
\fmfforce{1w,0.5h}{v6}
\fmf{plain}{v6,v5,v4,v3,v2,v1}
\fmf{boson,left}{v2,v3}
\fmf{boson,left}{v4,v5}
\fmfdot{v1,v2,v3,v4,v5,v6}
\end{fmfgraph}
\end{center}}
+
\parbox{28mm}{\begin{center}
\begin{fmfgraph}(25,8)
\setgras
\fmfforce{0w,0h}{v1}
\fmfforce{0.2w,0h}{v2}
\fmfforce{0.4w,11/16h}{v2b}
\fmfforce{0.8w,0h}{v3}
\fmfforce{0.6w,11/16h}{v3b}
\fmfforce{1w,0h}{v4}
\fmf{boson,left=0.33}{v2,v2b}
\fmf{boson,right=0.33}{v3,v3b}
\fmf{plain}{v4,v3,v2,v1}
\fmf{plain,right}{v2b,v3b,v2b}
\fmfdot{v1,v2,v3,v3b,v2b,v4}
\end{fmfgraph}
\end{center}}
\end{eqnarray*}

The number of Feynman diagrams at each order was calculated
in Ref.\cite{Riddell} (see also \cite{BrouderEPJC2}).

For the potential, we have the following diagrams
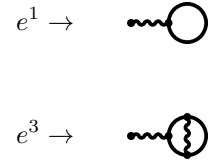
\begin{eqnarray*}
e^1 &\rightarrow&
\parbox{28mm}{\begin{center}
\begin{fmfgraph}(25,8)
\setgras
\fmfforce{0.2w,0.5h}{v2}
\fmfforce{0.4w,8/16h}{v2b}
\fmfforce{0.6w,8/16h}{v3b}
\fmf{boson}{v2,v2b}
\fmf{plain,right}{v2b,v3b,v2b}
\fmfdot{v2,v2b}
\end{fmfgraph}
\end{center}}
\\
e^3 &\rightarrow&
\parbox{28mm}{\begin{center}
\begin{fmfgraph}(25,8)
\setgras
\fmfforce{0.2w,0.5h}{v2}
\fmfforce{0.4w,8/16h}{v2b}
\fmfforce{0.5w,3/16h}{vxb}
\fmfforce{0.5w,13/16h}{vxc}
\fmfforce{0.6w,8/16h}{v3b}
\fmf{plain,right}{v2b,v3b,v2b}
\fmf{boson}{v2,v2b}
\fmf{boson}{vxb,vxc}
\fmfdot{v2,v2b,vxb,vxc}
\end{fmfgraph}
\end{center}}
\end{eqnarray*}

Each strong field diagram is the sum of an infinite number
of weak field diagrams, and each SCF diagram is the sum of an 
infinite number of strong field diagrams.

\section{The initial Green function \label{G0N}}

Once the functional equation is know, it remains to specify the
boundary conditions. For example, we can specify the Green function
without interaction. In the case of a static external field
$a_\mu({\mathbf{r}})$ or a static self-consistent field
$A_\mu({\mathbf{r}})$, this Green function is well-known and can
be written
\begin{eqnarray}
G_N^0(x,y) &=& i\theta(y^0-x^0)\sum_{E_n\le E_F} 
  \psi_n(x)\bar\psi_n(y)
\nonumber\\&&
             - i\theta(x^0-y^0)\sum_{E_n> E_F} 
  \psi_n(x)\bar\psi_n(y),
\label{GN0}
\end{eqnarray}
where $E_F$ is the Fermi energy, and where $\psi_n(x)$
is the solution of the Dirac equation at energy $E_n$.
Note that $G_N^0(x,y)=\hbar c S^a_N(x,y)$ 
or $G_N^0(x,y)=\hbar c S^A_N(x,y)$ where 
$S^a_N(x,y)$ and $S^A_N(x,y)$ are the Feynman electron Green functions
in the external potential $a_\mu(x)$ or $A_\mu(x)$ 
(with $N$ occupied bound states).
In the present section, we
want to discuss and justify this expression.

The notation follows standard textbooks on the Dirac equation 
\cite{Itzykson},\cite{Thaller}. The index $N$ 
means that $G_N^0(x,y)$ is the Green function for a total
charge of $Ne$. The Green function $G^0_0(x,y)$ represents
the vacuum in the presence of 
$a_\mu(\bfx)$
(or $A_\mu(\bfx)$).
This vacuum Green function is given by Eq.(\ref{GN0}) for
$E_F=0$.
The difference between $G_N^0(x,y)$ and $G^0_0(x,y)$ is
\begin{eqnarray*}
G_N^0(x,y)-G^0_0(x,y) &=& 
i\theta(y^0-x^0)\sum_{0<E_n\le E_F} \psi_n(x)\bar\psi_n(y)
\nonumber\\&&\hspace*{-20mm}
             + i\theta(x^0-y^0)\sum_{0<E_n\le E_F} \psi_n(x)\bar\psi_n(y),
\nonumber\\&=& i\sum_{0<E_n\le E_F} \psi_n(x)\bar\psi_n(y).
\end{eqnarray*}
The Fermi energy $E_F$ is determined by the condition
that the sum is over $N$ states:
\begin{eqnarray*}
\sum_{0<E_n\le E_F} 1 &=& N.
\end{eqnarray*}

\subsection{Early works}

To understand the origin of the Green function given
by Eq.(\ref{GN0}), it is useful to describe how
it appeared historically.

In the vacuum, Dirac \cite{Dirac33},\cite{Dirac34} assumed
that all the negative-energy states are filled.
This gives the density matrix
\begin{eqnarray*}
\rho(\bfx,\bfy)=\sum_{E_n<0} \psi^0_n(\bfx) {\psi^0_n}^\dagger(\bfy),
\end{eqnarray*}
where $\psi^0_n(\bfx)$ is the free solution of the Dirac
equation for energy $E_n$.
However, such a density matrix would be highly impractical,
since it would lead to an infinite charge density.
In a strike of genius, Dirac made the following observation
\cite{Dirac33},\cite{Dirac34}. Since
\begin{eqnarray*}
\sum_{E_n} \psi^0_n(\bfx) {\psi^0_n}^\dagger(\bfy) =\delta(\bfx-\bfy),
\end{eqnarray*}
the ``renormalized'' density matrix
\begin{eqnarray*}
\rho_0(\bfx,\bfy)&=&\rho(\bfx,\bfy)- \frac{1}{2}\delta(\bfx-\bfy)\\
  &=&\frac{1}{2}\sum_{E_n<0} \psi^0_n(\bfx) {\psi^0_n}^\dagger(\bfy)
    -\frac{1}{2}\sum_{E_n>0} \psi^0_n(\bfx) {\psi^0_n}^\dagger(\bfy)
\end{eqnarray*}
is much better behaved.\footnote{In fact, Dirac defined
$\rho_0(\bfx,\bfy)$ as the double of the right-hand side.
This was immediately corrected by Heisenberg \cite{Heisenberg}.}
In particular,
the charge density is
\begin{eqnarray*}
\rho_0(x)&=&\tr[\rho_0(x,x)]\\
  &=&\frac{1}{2}\sum_{E_n<0} |\psi^0_n(x)|^2
    -\frac{1}{2}\sum_{E_n>0} |\psi^0_n(x)|^2=0,
\end{eqnarray*}
where the last equality is due to charge conjugation symmetry.

In the presence of a static potential $a_\mu(\mathbf{x})$
or $A_\mu(\mathbf{x})$, 
and without bound states, the analogous density matrix is
\begin{eqnarray*}
\rho(x,y)
  &=&\frac{1}{2}\sum_{E_n<0} \psi_n(x) {\psi^0_n}^\dagger(\bfy)
    -\frac{1}{2}\sum_{E_n>0} \psi_n(x) {\psi^0_n}^\dagger(\bfy),
\end{eqnarray*}
where $\psi_n(x)$ is now the solution of the Dirac equation
with a potential $a_\mu(\mathbf{x})$ or $A_\mu(\mathbf{x})$ 
and energy $E_n$.
Of course, the charge conjugation symmetry is broken by
the external potential, but we still have that
\begin{eqnarray*}
\rho(x)&=&\tr\big(\rho(x,x)\big)=\tr\big(\rho(x,x)\big)-\rho_0(x)\\
 &=&\frac{1}{2}\sum_{E_n<0} (|\psi_n(x)|^2-|\psi^0_n(x)|^2)\\&&
    -\frac{1}{2}\sum_{E_n>0}(|\psi_n(x)|^2-|\psi^0_n(x)|^2)
\end{eqnarray*}
is less singular. The physical idea developed by
Dirac is to measure the charge density with respect to
charge density of the vacuum $\rho_0(x)$. The induced charge
density $\rho(x)\not=0$ in the presence of an
external potential was investigated in the thirties
\cite{Dirac33}, \cite{Dirac34}, \cite{Heisenberg},
\cite{Serber}, \cite{PauliRose}.
The potential corresponding to this induced charge
is called the Uehling potential \cite{Uehling}.
Physically, $\rho(x)\not=0$ is a reaction of the
vacuum which is polarized by the external potential.
The induced charge has observable consequences. For instance, its
effect on the energy levels of hydrogen atoms is well 
documented \cite{Mohr}. 
Computer programs are available
to evaluate the Uehling potential for general 
nuclear charge models \cite{Hnizdo}.  Expansions in
$(Z\alpha)^n$ where investigated in
\cite{Rinker}.

\subsection{Checking the Green function}
In this section, we check that Eq.(\ref{GN0})
gives the charge density proposed by Dirac.
In Eq.(\ref{GN0}) we write
\begin{eqnarray*}
\psi_n(x)=\exp(-iE_n x^0/c\hbar) \psi_n(\mathbf{x}),
\end{eqnarray*}
and we define the frequency dependent Green function
\begin{eqnarray*}
G_N^0(\bfx,\bfy;\omega) &=& \int_{-\infty}^\infty \dd t \,
\ee^{i\omega t} G_N^0(x,y), 
\end{eqnarray*}
where, on the right hand side, $x^0=y^0+ct$. Since the external
potential is time independent, the Green function depends only on $t$.
To determine $G_N^0(\bfx,\bfy;\omega)$ we 
use the distribution identity 
(Ref.\cite{Itzykson}, p.92)
\begin{eqnarray*}
\theta(x^0-y^0)=i\int_{-\infty}^\infty \frac{\dd \omega}{2\pi}
\frac{\exp{[-i\omega(x^0-y^0)]}}{\omega+i\epsilon}
\end{eqnarray*}
and we obtain 
\begin{eqnarray*}
G_N^0(\bfx,\bfy;\omega) &=& 
\sum_{E_n\le E_F} \frac{\psi_n(\bfx)\bar\psi_n(\bfy)}{\omega-E_n/\hbar-i\epsilon}
\\&&+
\sum_{E_n > E_F} \frac{\psi_n(\bfx)\bar\psi_n(\bfy)}{\omega-E_n/\hbar+i\epsilon}.
\end{eqnarray*}
This is the standard expression for the non-interacting electron Green function
in the relativistic case (see Ref.\cite{Mohr} p. 32 for $E_F=0$) and as well as in
the nonrelativistic case (up to a factor $\gamma^0$,
see Ref. \cite{Fetter}, p.124).

The induced current is 
$-ie \tr[\gamma^0 G_N^0(x,x)]$. But, this
formal expression is infinite and must be regularized.
The limit involved in the expression $G_N^0(x,x)$
is discussed in detail in section \ref{currentsec}:
\begin{eqnarray*}
G_N^0(x,x) &=& 
\frac{1}{2} \lim_{\zbf\rightarrow 0} 
\lim_{z^0\rightarrow 0^+} 
\big(G_N^0(x+z,x-z)
\\&&
+G_N^0(x-z,x+z)\big).
\end{eqnarray*}
This definition agrees with experiments in nuclear physics
\cite{Greiner} and atomic physics (Ref.\cite{Mohr} p. 275).

With this definition, we obtain 
\begin{eqnarray*}
G_N^0(x,x) &=& \frac{i}{2} \Big(\sum_{E_n\le E_F} \psi_n(\bfx)\bar\psi_n(\bfx)
-\sum_{E_n > E_F} \psi_n(\bfx)\bar\psi_n(\bfx)\Big).
\end{eqnarray*}
In particular, we obtain the charge density:
\begin{eqnarray*}
\rho(\bfx) &=& J^0(\bfx) = 
-ie \tr[\gamma^0 G_N^0(x,x)] \\
&=& \frac{e}{2} \Big(\sum_{E_n\le E_F} |\psi_n(\bfx)|^2 
 - \sum_{E_n > E_F}|\psi_n(\bfx)|^2\Big).
\end{eqnarray*}

This is indeed the density obtained by Dirac.
One might wonder how this charge density is related
to the many-body charge density (or the non-relativistic one), 
which is obtained by
summing over the occupied states with positive energy.
The answer is that we must
subtract the contribution of the Green function
$G^0_0(x,x)$ (for which $E_F=0$).
Then we obtain
\begin{eqnarray*}
-ie\tr[\gamma^0 \big(G_N^0(x,x)-G^0_0(x,x)\big)] &=&
e\hspace*{-2mm} \sum_{0<E_n\le E_F} \hspace*{-2mm}|\psi_n(\bfx)|^2.
\end{eqnarray*}
The right hand side is the familiar expression 
for $\rho(\bfx)$.
It is not strictly correct, since
it neglects the effect of $G^0_0(x,x)$, which describes the
polarisation of the vacuum by the external potential (i.e.
the Uehling potential, see Ref.\cite{Itzykson} p. 327
or Ref.\cite{Mohr} p.265). It might seem strange that the
QED and the many-body densities are different, since they
are given by the same equation
\begin{eqnarray*}
\rho(\bfx) &=& -\frac{ie}{2\pi} \int_{-\infty}^\infty \dd\omega\,
\tr[\gamma^0 G_N^0(\bfx,\bfx;\omega)].
\end{eqnarray*}
The reason for this
discrepancy is that the expression for $\rho(\bfx)$ is divergent,
and the difference comes from a different renormalization of it.
In the many-body case, an innocent-looking convergence
factor $\ee^{i\omega\eta}$ is introduced in the integral
over $\omega$ (see Ref.\cite{Fetter} p.124), which selects 
precisely the states $E_n\le E_F$ (the states $E_n<0$ are
discarded from the start).
On the other hand, in the QED case, the integration contour
is below the real axis from $-\infty$ to $E_F$ and above
it from $E_F$ to $\infty$.

Several approximations are possible (no-sea, no-pair). They are
discussed in Refs.\cite{Engel3}, \cite{Facco}.

\subsection{Subsequent works}

This idea of a Dirac sea in an external potential
has received many experimental and theoretical confirmations.
The first and strongest one being the discovery of the 
positron.

Since then, the Dirac sea and Eq.(\ref{GN0}) were used as a basis 
for strong field electrodynamics
\cite{Greiner}, \cite{Reinhard}, \cite{Rafelski}, \cite{Plunien}, 
for thermal field theory
\cite{Bowers2}, \cite{Landsman},
\cite{LeBellac}, \cite{Borisov}, 
and for relativistic quantum many-body
 theory in Riemannian spacetime \cite{Bowers}.

The concept of a Dirac sea is sometimes considered to
be rather out of date. However, according to Jackiw
\cite{Jackiw2}, it is still the best physical picture to understand
striking phenomena such as chiral anomalies \cite{Bertlmann}
and fractional fermion numbers \cite{Niemi}.

The case when the Fermi energy is degenerate
will not be considered in this paper.

\section{The induced current \label{currentsec}}
It is important to derive the proper expression of the
four-current as a function of the electron Green function.
As a field operator, the current must be self-adjoint
(it is measurable), conserved, gauge invariant and 
should change sign under charge conjugation.

\subsection{Definition}

We follow Refs.\cite{Pauli}, \cite{SchwingerGauge}, \cite{Dosch}
and define the bilocal field operator
\begin{eqnarray*}
\Jbf^\mu(x,y) &=& \frac{e}{2} \sum_{\alpha\beta}
\gamma^\mu_{\alpha\beta}
\big[\bar\psibf_{\alpha}(x) \psibf_{\beta}(y)
    -\psibf_{\beta}(x)\bar\psibf_{\alpha}(y)\big].
\end{eqnarray*}

When this operator is evaluated at the ground
state of the system, its value can be expressed
in terms of the electron Green function.

The Stueckelberg-Feynman Green functions that we use
in the present paper are defined from the time-ordering
operator (\cite{Itzykson} p.150)
\begin{eqnarray*}
T\psibf_\alpha(x)\bar\psibf_\beta(y) &=&
\theta(x^0-y^0) \psibf_\alpha(x)\bar\psibf_\beta(y)
\\&&
- \theta(y^0-x^0) \bar\psibf_\beta(y)\psibf_\alpha(x)
\end{eqnarray*}
by
\begin{eqnarray}
\langle T\psibf_\alpha(x)\bar\psibf_\beta(y) \rangle &=&
i\hbar c S_{\alpha\beta}(x,y).\label{S=Tpsi}
\end{eqnarray}
We consider here the interacting fields and not the
free fields as in Ref.\cite{Itzykson} p.272.
Therefore, $S(x,y)$ is the full electron Green function.

To define the current we follow Schwinger's prescription
\cite{SchwingerGauge}. Let $z$ be a four-vector with
$z^0>0$, the four-current is 
\begin{eqnarray}
J^\mu(x) &=& -ie\hbar c 
\tr\big[\gamma^\mu S(x,x)], \label{defjmu}
\end{eqnarray}
where $S(x,x)$ is defined by the symmetric limit
\begin{eqnarray}
S(x,x) &=& \frac{1}{2} \lim_{\zbf\rightarrow 0} 
\lim_{z^0\rightarrow 0^+} 
\big(S(x+z,x-z)
\nonumber\\&&
+S(x-z,x+z)\big).\label{deflim}
\end{eqnarray}
More precisely, we start from a non-zero 3-vector
$\zbf$ (i.e. $|\zbf|\not=0$) and a positive
time $z^0$. We first make $z^0\rightarrow 0$
and then $|\zbf|\rightarrow 0$.
Because of the symmetric limit we obtain
the same result with $z^0<0$
(just take $z\rightarrow -z$ in the above
definition)
\begin{eqnarray*}
S(x,x) &=& \frac{1}{2} \lim_{\zbf\rightarrow 0}
\lim_{z^0\rightarrow 0^-}
\big(S(x+z,x-z)
\\&&
+S(x-z,x+z)\big).
\end{eqnarray*}

From definition (\ref{S=Tpsi}), for $z^0>0$ we have
\begin{eqnarray}
S_{\alpha\beta}(x+z,x-z) &=&
-\frac{i}{\hbar c} 
\langle \psibf_\alpha(x+z)\bar\psibf_\beta(x-z) \rangle,
\label{Sxplusz}\\
S_{\alpha\beta}(x-z,x+z) &=&
\frac{i}{\hbar c} 
\langle \bar\psibf_\beta(x+z)\psibf_\alpha(x-z) \rangle.
\label{Sxmoinsz}
\end{eqnarray}
When we compare this with the definition of the
bilocal current field operator, we obtain
\begin{eqnarray*}
J^\mu(x) &=& \lim_{\zbf\rightarrow 0} 
\lim_{z^0\rightarrow 0^+} 
\langle \Jbf^\mu(x+z,x-z) \rangle.
\end{eqnarray*}
On the other hand, if $z^0<0$ 
\begin{eqnarray*}
J^\mu(x) &=& \lim_{\zbf\rightarrow 0}
\lim_{z^0\rightarrow 0^-}
\langle \Jbf^\mu(x-z,x+z)\rangle.
\end{eqnarray*}
Thus, when the definition of 
$J^\mu(x)$ is written in terms of
the bilocal operator, the first
argument of the bilocal operator is always
later than $x$. It is not clear
that the current defined by Eq.(\ref{defjmu}) 
is real, as a measurable
quantity should be.

Now we prove that the current 
is real and transforms properly under
charge conjugation.

\subsection{Self-adjointness}
Using the definition $\bar \psibf=\psibf^\dagger\gamma^0$,
the bilocal operator can be rewritten
\begin{eqnarray*}
\Jbf^\mu(x,y) &=& \frac{e}{2} 
\big[\psibf(x)^\dagger\gamma^0\gamma^\mu \psibf(y)
    -\psibf(x)(\gamma^0\gamma^\mu)^T\psibf(y)^\dagger\big],
\end{eqnarray*}
where $A^T$ is the transpose of $A$.
From the identity $(\gamma^0\gamma^\mu)^\dagger=\gamma^0\gamma^\mu$
(Ref.\cite{Itzykson} p.693)
we obtain immediately
\begin{eqnarray}
\Jbf^\mu(x,y)^\dagger=\Jbf^\mu(y,x). \label{selfadjointJ}
\end{eqnarray}

Now we can make an important remark concerning 
the bilocal current. If $x$ and $y$ are separated
by a space-like interval (i.e. if $(x-y)\cdot(x-y)<0$),
then $\bar\psibf(x)$ and $\psibf(y)$ anticommute.
This property is called causality and
means that two events that are
too far to be linked by a light-ray are independent.
Causality is a basis of all axiomatic approaches to
quantum field theory \cite{Haag}.

Thus, if $(x-y)\cdot(x-y)<0$ we have
\begin{eqnarray*}
\Jbf^\mu(x,y) &=& \frac{e}{2} \sum_{\alpha\beta}
\gamma^\mu_{\alpha\beta}
\big(-\psibf_{\beta}(y)\bar\psibf_{\alpha}(x)
    +\bar\psibf_{\alpha}(y)\psibf_{\beta}(x)\big)
\\&=& \Jbf^\mu(y,x).
\end{eqnarray*}
Combined with Eq.(\ref{selfadjointJ}), this means
that the bilocal current is self-adjoint for
space-like separated points.

The symmetric definition (\ref{deflim}) of the 
current in terms of the
Green function was chosen such that 
first $z^0\rightarrow 0^+$ with
$|\zbf|\not=0$. This is to ensure that,
for $z^0$ small enough (i.e. $z^0 < |\zbf|$), 
the interval $z$ becomes space-like and the current
is real. 

Moreover, if $x$ and $y$ are space-like separated,
we can rewrite the bilocal current in terms
of field commutators:
\begin{eqnarray}
\Jbf^\mu(x,y) &=& \frac{\Jbf^\mu(x,y)+\Jbf^\mu(y,x)}{2}
\nonumber\\&=&\frac{e}{4} \sum_{\alpha\beta}
\gamma^\mu_{\alpha\beta}
\big([\bar\psibf_{\alpha}(x),\psibf_{\beta}(y)]
    +[\bar\psibf_{\alpha}(y),\psibf_{\beta}(x)]\big).
\nonumber\\&&\label{Jcommutator}
\end{eqnarray}

The fact that the current $J^\mu(x)$ defined
by Eq.(\ref{defjmu}) is real can also be obtained
directly from a property of $S(x,y)$.
If $z$ is a space-like vector, causality ensures that
\begin{eqnarray*}
{\big(\psi_\alpha(x+z)\psi^\dagger_\beta(x-z)\big)}^\dagger
&=&
\psi_\beta(x-z)\psi^\dagger_\alpha(x+z)
\\&=&
-\psi^\dagger_\alpha(x+z)\psi_\beta(x-z).
\end{eqnarray*}
Now Eq.(\ref{Sxplusz}) gives
\begin{eqnarray*}
\langle \psi_\alpha(x+z)\psi^\dagger_\beta(x-z)\rangle &=& 
i\hbar c \sum_\lambda S_{\alpha\lambda}(x+z,x-z)\gamma^0_{\lambda\beta},
\end{eqnarray*}
and Eq.(\ref{Sxmoinsz}) gives
\begin{eqnarray*}
\langle \psi^\dagger_\alpha(x+z)\psi_\beta(x-z)\rangle &=& 
-i\hbar c \sum_\mu S_{\beta\mu}(x+z,x-z)\gamma^0_{\mu\alpha}.
\end{eqnarray*}
Thus, we obtain, for space-like $z$
\begin{eqnarray}
S^\dagger(x+z,x-z) &=& - \gamma^0 S(x-z,x+z) \gamma^0.
\label{Sdagger}
\end{eqnarray}
Here the dagger operator acts only on the spin-variables:
${S^\dagger}_{\alpha\beta}(x+z,x-z)=S^*_{\beta\alpha}(x+z,x-z)$.
From Eq.(\ref{Sdagger}) and the
identity $\gamma^0 \gamma^\mu \gamma^0 = {\gamma^\mu}^\dagger$
we can check that ${J^\mu(x)}^*=J^\mu(x)$.

\subsection{Charge conjugation}
Let $\Ccal$ be the charge conjugation operator acting on fields and
$C=i\gamma^2\gamma^0$ the charge conjugation matrix.
We have the identities (\cite{Itzykson} 
p.152\footnote{Their misprinted definition of $C$ is corrected on 
page 693 of \cite{Itzykson}.}
 and \cite{Peskin} p.70)
\begin{eqnarray*}
\Ccal \psibf(x;e) \Ccal^\dagger &=& C \bar\psibf(x;-e), \\
\Ccal \bar\psibf(x;e) \Ccal^\dagger &=& \psibf(x;-e)C.
\end{eqnarray*}
In the presence of an external field, the sign of the
charge in the field operators is reversed under charge conjugation
(\cite{Pauli} p.19). 

The action of the charge operator on the bilocal current is
\begin{eqnarray*}
\Ccal \Jbf^\mu(x,y;e) \Ccal^\dagger &=&
\frac{e}{2} \big(\psibf(x;-e) C\gamma^\mu C \bar\psibf(y;-e)
\\&&
               -\bar\psibf(x;-e) (C\gamma^\mu C)^T \psibf(y;-e) \big).
\end{eqnarray*}
From the identities (\cite{Itzykson} p.693)
$C\gamma^\mu C^\dagger=-{\gamma^\mu}^T$ and
$C^T=C^\dagger=C^{-1}=-C$ we obtain
\begin{eqnarray*}
\Ccal \Jbf^\mu(x,y;e) \Ccal^\dagger &=& -\Jbf^\mu(x,y;-e).
\end{eqnarray*}
Therefore, the sign of the bilocal current is reversed
under charge conjugation.

Now we investigate the behaviour of the current
under charge conjugation, when it is defined
in terms of the Green function. To do this, we need an
equation for the Green function of the charge conjugated
problem. We obtain it by transposing both
sides of Eq.(\ref{anothereq}). This transposition is
meant for the 4x4 matrices only:
${S^T}_{ss'}(x,y)=S_{s's}(x,y)$.
Then we multiply both sides of the equation by 
the charge conjugation matrix $C$ on the
left and $C^{-1}$ on the right. We use the property
$C{\gamma^\mu}^T C^{-1}=-\gamma^\mu$ and we obtain
\begin{eqnarray*}
\delta(x-y) &=&
(i\hbar c \gamma\cdot\partial -mc^2+e\gamma\cdot A(x))S^c(x,y) \nonumber\\&&
 +ie\hbar c\gamma_\mu
\frac{\delta S^c(x,y)}{\delta j_\mu(x)},
\end{eqnarray*}
where we have defined
$S^c(x,y)=C S(y,x)^T C^{-1}$. If we compare this equation
with the second basic equation (\ref{secondbasic}),
we see that they become identical if we make the transformation
$e\rightarrow -e$, or equivalently
$A^\mu\rightarrow -A^\mu$ and $j^\mu\rightarrow -j^\mu$.
But, using Eq.(\ref{eqA}) and the charge conjugation
property of the induced current $J^\mu(x)$, we
see that $j^\mu\rightarrow -j^\mu$ implies
$A^\mu\rightarrow -A^\mu$.
In other words, we have the symmetry
\begin{eqnarray*}
S^c(x,y) &=& S(x,y;-j) = C S(y,x;j)^T C^{-1}.
\end{eqnarray*}
Therefore
\begin{eqnarray*}
\tr[\gamma^\mu S(x-z,x+z;j)] &=&
\tr[C\gamma^\mu C^{-1} S(x+z,x-z;-j)^T]
\\&=&
-\tr[{\gamma^\mu}^T S(x+z,x-z;-j)^T]
\\&=& -\tr[\gamma^\mu S(x+z,x-z;-j)].
\end{eqnarray*}
And the current can be rewritten
as the antisymmetric limit
\begin{eqnarray*}
S(x,x;j) &=& \frac{1}{2} \lim_{\zbf\rightarrow 0} 
\lim_{z^0\rightarrow 0^+} 
\big(S(x+z,x-z;j)
\nonumber\\&&
-S(x+z,x-z;-j)\big).
\end{eqnarray*}
This equation shows that
the current is an odd function of the external
current: $J_\mu(x;j)=-J_\mu(x;-j)$.
Changing the sign of $j$ amounts to changing
the sign of $A$, and we can also write
$J_\mu(x;A)=-J_\mu(x;-A)$: the current is
an odd functions of $A$. Now, since the
electron loops of QED are generated by
the current, and since the $x$ in
$S(x,x;A)$ is linked to another photon
propagator, we see that the electron loops
with an odd number of photon lines 
are zero.
This is a 
version of Furry's theorem for QED with
external field.
The relevance of charge conjugation for the
QED current was already noticed by Kramers in
1937 \cite{Kramers}. Related results can
be found in Ref.\cite{Greiner}, sections
4.2, 9.4 and 15.1.

\subsection{Charge conservation}
Charge conservation is described by the equation
$\partial_\mu J^\mu(x)=0$. How does this translate
for our bilocal operator? We want a real and symmetric
expression for $\partial_\mu J^\mu(x)=0$. 
Why symmetric? If a bilocal operator $F(x,y)$ is 
antisymmetric (i.e. $F(x,y)=-F(y,x)$),
then its value at $x=y$ is zero, except
at divergences, such as $(x-y)/(x-y)^2$.
Since the regular part of $F(x,y)$ is zero
on the diagonal, the use of a symmetric operator 
eliminates the antisymmetric divergences
without changing the value of the 
regular operator on the diagonal.

An obvious candidate is
$(\partial_\mu^x+\partial_\mu^y) \Jbf^\mu(x,y)/2$.
From the field equations \cite{Greiner} p.195
\begin{eqnarray*}
(i\hbar c\partial_\mu - e A_\mu)\gamma^\mu\psibf -m c^2\psibf  &=& 0, \\
(i\hbar c\partial_\mu + e A_\mu)\bar\psibf\gamma^\mu +m c^2\bar\psibf  &=& 0,
\end{eqnarray*}
we check that $(\partial_\mu^x+\partial_\mu^y) \Jbf^\mu(x,y)=0$.

In terms of the electron Green function, this becomes
\begin{eqnarray*}
-\frac{ie\hbar c}{4} \partial_\mu^x
\tr[\gamma^\mu S(x+z,x-z)+
\gamma^\mu S(x+z,x-z)],
\end{eqnarray*}
where we have used
$\partial_\mu^{x+z}+\partial_\mu^{x-z}=\partial_\mu^x$.
and where $z$ is space-like.
Charge conservation can then be directly verified
from Eqs.(\ref{secondbasic}) and (\ref{anothereq}).


\subsection{Mass-reversal symmetry}
The concept of mass-reversal symmetry was discussed
by Peaslee \cite{Peaslee} and Tiomno \cite{Tiomno}.
For massless particles, mass-reversal symmetry is known
as chiral symmetry.

The mass-reversed fermion field $\psibf(-m)$ satisfies the
same equations as the usual fermion field $\psibf(m)$, but
with a reversed mass:
\begin{eqnarray*}
(i\hbar c\gamma\cdot\partial - e \gamma\cdot A +m c^2)\psibf(-m)  &=& 0.
\end{eqnarray*}
Since $\gamma^\mu\gamma_5=-\gamma_5\gamma^\mu$ we can take
$\psibf(x;-m)=\gamma^5\psibf(x;m)$ (up to a phase factor $\eta_m$
which disappears in the Green function).
To see how the Green function transforms under mass reversal, 
we multiply both sides of Eq.(\ref{secondbasic}) by $\gamma_5$
on the left and on the right. The commutation rules between
$\gamma_5$ and $\gamma^\mu$ yield
\begin{eqnarray}
\delta(x-y) &=&
(-i\hbar c \gamma\cdot\partial -mc^2+e\gamma\cdot A(x)) 
\gamma_5 S(x,y;m)\gamma_5 \nonumber\\&&
 +ie\hbar c\gamma_\mu
\gamma_5 \frac{\delta S(x,y;m)}{\delta j_\mu(x)}\gamma_5 ,
\end{eqnarray}
which is the equation for $-S(x,y;-m)$.
Therefore,
\begin{eqnarray*}
S(x,y;-m)=-\gamma_5 S(x,y;m)\gamma_5,
\end{eqnarray*}
and the mass-reversed current is
\begin{eqnarray*}
J^\mu(x;-m) &=& -ie\hbar c \tr[\gamma^\mu S(x,x;-m)]
\\
&=&ie\hbar c \tr[\gamma_5\gamma^\mu\gamma_5 S(x,x;m)] = J^\mu(x;m).
\end{eqnarray*}
Thus, the induced current is even under mass-conjugation, and
the vector potential also, as expected \cite{Tiomno}.

In other words, the induced current is an even function of
$m$.

\subsection{Gauge invariance}
The Lagrangian of QED is invariant under the
gauge transformations (\cite{Itzykson} p.64)
\begin{eqnarray*}
\psibf^\Lambda(x) &=& \ee^{-ie\Lambda(x)} \psibf(x),
\\
\quad \Abf^\Lambda_\mu(x) &=& \Abf_\mu(x) + \hbar c\partial_\mu \Lambda(x).
\end{eqnarray*}
After a gauge transformation, the bilocal current
for space-like separated points Eq.(\ref{Jcommutator})
becomes
\begin{eqnarray}
\Jbf^\Lambda(x,y) &=& \cos(e\Lambda(x)-e\Lambda(y)) \Jbf(x,y)
\nonumber\\&&
+\frac{ie}{4}\sin(e\Lambda(x)-e\Lambda(y))
\nonumber\\&&\hspace*{-10mm}\times
\sum_{\alpha\beta}
\gamma^\mu_{\alpha\beta}
\big([\bar\psibf_{\alpha}(x),\psibf_{\beta}(y)]
    -[\bar\psibf_{\alpha}(y),\psibf_{\beta}(x)]\big).
\label{Jprime}
\end{eqnarray}
In terms of the electron Green function, we must
consider the transformed Green functions
\begin{eqnarray}
S^\Lambda(x,y)&=& \ee^{-ie\Lambda(x)+ie\Lambda(y)} S(x,y).
\label{SLambda}
\end{eqnarray}
Hence,
\begin{eqnarray*}
S^\Lambda(x+z,x-z)+S^\Lambda(x-z,x+z) &=&
\\&&\hspace*{-60mm}
\cos(e\Lambda(x)-e\Lambda(y))
\big(S(x+z,x-z)+S(x-z,x+z)\big)
\\&&\hspace*{-60mm}
-i\sin(e\Lambda(x)-e\Lambda(y))
\big(S(x+z,x-z)-S(x-z,x+z)\big).
\end{eqnarray*}
A priori, the singularities
of the Green function are not weak enough to 
make the second term tend to zero when
$z\rightarrow0$. Therefore, our definition of the
current is not obviously gauge invariant.

This problem can be solved by using a Green function
multiplied by Dirac's phase \cite{Dirac34}
\begin{eqnarray*}
S'(x+z,x-z) &=&
\exp\big[\frac{ie}{\hbar c} 
\int_{-1}^1 \dd\lambda\, z^\mu A_\mu(x+\lambda z)\big]
\\&&\times S(x+z,x-z).
\end{eqnarray*}
A generalization of this phase was used
recently for the renormalization of
infrared divergences in QED
\cite{BaganI}, \cite{BaganII}.

The modified Green function
$S'(x+z,x-z)$ is gauge invariant because,
according to Eq.(\ref{ndim}) in the appendix,
\begin{eqnarray*}
\int_{-1}^1 \dd\lambda \partial_\mu\Lambda(x+\lambda z)
z^\mu &=& \Lambda(x+z)-\Lambda(x-z).
\end{eqnarray*}
The right hand side compensates for the
gauge transformation phase in Eq.(\ref{SLambda}),
so that ${S'}^\Lambda(x,y)=S'(x,y)$.

If we substitute $S'(x+z,x-z)$
for $S(x+z,x-z)$ in the definition
of the current (\ref{defjmu}), we obtain 
a current which is real, odd under
charge conjugation and gauge invariant,
but charge conservation becomes 
doubtful. 
Thus, we stick to the 
original definition
Eq.(\ref{defjmu}).

\section{Renormalization}

The renormalization of QED in the presence of an external field
does not seem to have been treated in detail.
Several attempts exist 
\cite{Engel}, \cite{Engel4}, 
\cite{Reinhard}, \cite{Dosch},
but none of them was fully completed.
We start now a renormalization of QED in an external
field which closely follows the renormalization
of vacuum QED.
The first step is to determine Dyson's relations,
which are non perturbative expressions that link
renormalized and non renormalized propagators and
fields.
Then we give a non perturbative relation between
the renormalized potential and the renormalized
vacuum polarization.
Finally, we describe the renormalization
rules that enable us to obtain finite quantities
for any strong field or SCF Feynman diagram.

\subsection{Dyson relations for strong field QED}
The first point is to determine the Dyson relations between 
renormalized
and unrenormalized propagators. This relation is given by 
Sterman\cite{Sterman} for scalar fields and
we derive it for QED.
If we denote the renormalized quantities with a overlined symbol,
in the presence of an external potential $a_\mu(x)$, 
Dyson's relations are
\cite{Dyson}
\begin{eqnarray*}
S(x,y;a;e_0,m_0) &=& Z_2 \barS(x,y; \sqrt{Z_3} a;e,m),\\
D(x,y;a;e_0,m_0) &=& Z_3 \barD(x,y; \sqrt{Z_3} a;e,m),\\
A(x;a;e_0,m_0) &=& \sqrt{Z_3} \barA(x; \sqrt{Z_3} a;e,m).
\end{eqnarray*}

To derive these relations, we expand $S$ over $a$:
\begin{eqnarray*}
S(x,y;a) &=& S(x,y; 0) + \int \dd z_1 \frac{\delta S(x,y;0)}{\delta a_\mu(z_1)}  
  a_\mu(z_1) +\cdots
\end{eqnarray*}
The Dyson relation for the vacuum propagator $ S(x,y; 0)$ is
$ S(x,y; 0) = Z_2 \bar  S(x,y; 0)$. For the other terms, the functional
derivative with respect to the external potential brings a factor 
$A(x)$ in the path integral. Therefore,
\begin{eqnarray*}
 \frac{\delta ^nS(x,y;0)}{\delta a^n}= Z_2 Z_3^{n/2}
    \frac{\delta^n \barS(x,y;0)}{\delta a^n}.
\end{eqnarray*}
Hence,
\begin{eqnarray*}
S(x,y;a) &=& Z_2 \barS(x,y; 0) + Z_2 \int \dd z_1 
   \frac{\delta\barS(x,y;0)}{\delta a_\mu(z_1)}  
   \sqrt{Z_3}a_\mu(z_1)\\&& +\cdots\\
&=& Z_2 \barS(x,y;\sqrt{Z_3} a).
\end{eqnarray*}
Or, more precisely,
\begin{eqnarray*}
S(x,y;a;e_0,m_0) &=& Z_2 \barS(x,y;\sqrt{Z_3} a;e,m).
\end{eqnarray*}

Because of this relation, we also define the renormalized
external potential as $\bara_\mu(x)=\sqrt{Z_3} a_\mu(x)$.
It may seem strange that the external potential $a_\mu(x)$
and the full potential $A_\mu(x)$ are not renormalized
with the same formula. However, we must recall that
$a_\mu(x)$ was defined in terms of an external
current $j_\mu(x)$, which was used to generate
Green functions by functional derivatives.
The bare Green functions are generated
by $A^\mu(x) j_\mu(x)$ and the
renormalized ones by
$\barA^\mu(x) \barj_\mu(x)$.
Since we want these two terms to be equal,
the relation $\sqrt{Z_3}\barA_\mu(x)=A_\mu(x)$
implies $\barj_\mu(x)=\sqrt{Z_3}j_\mu(x)$
(see Refs.\cite{Sterman} p. 297 and \cite{Ticciati} p.288).
More physically, we can say that an external field
is made by preparing a certain density of (charged) matter
$n(\bfx)$.  Before renormalization, this density corresponds
to a charge density $j^0(\bfx)=e_0 n(\bfx)$. The
renormalization
modifies the charge, but not the density of matter. Thus
the renormalized charge density is now
$\barj^0(\bfx)=e n(\bfx) =\sqrt{Z_3} j^0(\bfx)$,
and the corresponding relation for the external
potential is again $\bara^\mu(x)=\sqrt{Z_3} a^\mu(x)$.

In terms of the renormalized propagators, the functional
equation becomes (using $e_0=e/\sqrt{Z_3}$ and
$m_0=m+\delta m$)
\begin{eqnarray*}
(i\hbar c \gamma\cdot\partial -mc^2-\delta mc^2-
e\gamma\cdot \barA(x))
  Z_2\barS(x,y) &=& \delta(x-y)\\
&&\hspace*{-60mm} +ie\hbar c Z_2 \gamma^\mu
\int ds \frac{\delta \barS(x,y)}{\delta \bara_{\lambda}(s)}
  D^0_{\lambda\mu}(s,x).
\end{eqnarray*}
and

\begin{eqnarray*}
\sqrt{Z_3}\barA_\mu(x) &=& \bara_\mu(x)/\sqrt{Z_3}
 -i\frac{e}{\sqrt{Z_3}}\hbar c Z_2\\&&\times 
\int ds D^0_{\mu\nu}(x,y)
   \tr[\gamma^\nu \barS(y,y)],
\end{eqnarray*}
or
\begin{eqnarray}
\barA_\mu(x) &=& \frac{\bara_\mu(x)}{Z_3}
 -i\frac{e Z_2}{Z_3}\hbar c \int ds D^0_{\mu\nu}(x,y)
   \tr[\gamma^\nu \barS(y,y)],
\nonumber\\&&\label{AZ2Z3}
\end{eqnarray}

\subsection{Dyson relations for SCF-QED}
Notice that, in the case of SCF-QED, Dyson's relations
are much simpler, because $e_0 A=e\barA$:
\begin{eqnarray}
S(x,y;e_0 A;e_0,m_0) &=& Z_2 \barS(x,y; e\barA;e,m),
\label{DysonrelS}\\
D(x,y;e_0 A;e_0,m_0) &=& Z_3 \barD(x,y; e\barA;e,m),\\
A(x; e_0 A;e_0,m_0) &=& \sqrt{Z_3} \barA(x; e\barA;e,m).
\end{eqnarray}

The main difference between vacuum QED and SCF-QED
is the renormalization of the induced vector potential.
The total bare potential is 
\begin{eqnarray}
A_\mu(x) &=& a_\mu(x)
 -ie_0\hbar c \int \dd s D^0_{\mu\nu}(x,s)
 \tr[\gamma^\nu S(s,s)],\,\,
\label{AparS}
\end{eqnarray}
and the problem is the renormalization of 
$\tr[\gamma^\nu S(s,s)]$. To indicate that
the electron Green function is calculated in the presence
of the full potential $A_\mu(x)$ we denote it
by $S(x,y;A)$.

Dyson's relation (\ref{DysonrelS}) gives us
\begin{eqnarray*}
\tr[\gamma^\nu S(s,s;A)] &=& Z_2 
\tr[\gamma^\nu \barS(s,s;\barA)].
\end{eqnarray*}
We rewrite the right hand side using Eq.(\ref{appendiceIntegral}):
\begin{eqnarray*}
\tr[\gamma^\nu \barS(s,s;\barA)] &=&
\tr[\gamma^\nu \barS(s,s;0)] 
+\int_0^1\dd\lambda
\int \dd x \\&&\hspace*{-20mm}\times
\tr\Big[\gamma^\nu \frac{\delta \barS(s,s;\lambda \barA)}
{\delta \barA^{\mu}(x)}\Big] \barA^{\mu}(x).
\end{eqnarray*}

Now $-ie\hbar c\tr[\gamma^\nu \barS(s,s;0)]$ is the
current in the presence of a zero external field $\barA=0$.
Thus, it is the vacuum current, which is zero
(more precisely, $\tr[\gamma^\nu \barS(s,s;0)]$ is 
made of fermion loops
with one external photon line, which are zero by Furry's theorem).

The second term is also well known from the renormalization
of the vacuum polarization (see e.g. Eq.(22) of Ref.\cite{BFI})
\begin{eqnarray}
-ie\hbar c Z_2 \tr \Big[\gamma_\nu
\frac{\delta\barS(s,s;\lambda\barA)}{\delta \barA^\mu(y)}\Big]
&=&\nonumber\\&&\hspace*{-40mm}
\barPi_{\nu\mu}(s,y;\lambda\barA)
+(Z_3-1) {D^0}^{-1}_{\mu\nu}(s,y).
\label{Pi}
\end{eqnarray}
Thus,  we obtain
\begin{eqnarray*}
-ie\hbar c Z_2 \int \dd y \tr [\gamma_\nu
\barS(s,s;\barA)]
&=& 
\\&&\hspace*{-20mm}
\int_0^1 \dd\lambda
\int \dd y \barPi_{\nu\mu}(s,y;\lambda\barA) \barA^\mu(y)
\\&&\hspace*{-20mm}
+(Z_3-1) \int \dd y {D^0}^{-1}_{\mu\nu}(s,y) \barA^\mu(y) .
\end{eqnarray*}

After this reorganization of the induced current, 
Eq.(\ref{AparS}) becomes
\begin{eqnarray*}
A_\mu(x) &=& a_\mu(x)
+ \frac{Z_3-1}{\sqrt{Z_3}} 
 \barA_\mu(x)
\\&&
+ \frac{1}{\sqrt{Z_3}} 
 \int \dd s \dd y \dd\lambda D^0_{\mu\nu}(x,s)
 \barPi^{\nu\rho}(s,y;\lambda\barA) \barA_\rho(y).
\end{eqnarray*}
Now we know that the relation between the bare and
renormalized potentials is $A_\mu(x)=\sqrt{Z_3}\barA_\mu(x)$.
This yields
\begin{eqnarray}
\barA_\mu(x) &=& {\bara}_\mu(x)
+ \int_0^1\dd\lambda\int \dd s \dd y D^0_{\mu\nu}(x,s)
\nonumber\\&&\times
 \barPi^{\nu\rho}(s,y;\lambda\barA) \barA_\rho(y),
\label{Aren=D0PiA}
\end{eqnarray}
where the true external potential $a_\mu(x)$
is renormalized by $\bara_\mu(x)=\sqrt{Z_3}a_\mu(x)$.

It should be noticed that, in all our manipulations,
the renormalization factors $Z_2$, $Z_3$ and  $\delta m$ where
taken at $\barA_\mu=0$. In other words, the renormalization
factors of QED with external field are the same as
the renormalization factors of vacuum QED. More
precisely, the vacuum renormalization factors removes the
divergences of QED in external field. However, 
renormalization conditions may introduce finite differences
between the renormalization factors of vacuum QED and
QED in an external field.

From Eq. (\ref{Aren=D0PiA}), it can be checked that
$\barPi$ is indeed the renormalized vacuum polarization.
If we write $X_\mu(x)$ for the second term on the right hand
side of Eq. (\ref{Aren=D0PiA}) we find
\begin{eqnarray}
\barD_{\mu\nu}(x,y) &=& \frac{\delta \barA_\mu(x)}{\delta \barj^\nu(y)}
=
\frac{\delta \bara_\mu(x)}{\delta \barj^\nu(y)}+
\frac{\delta X_\mu(x)}{\delta \barj^\nu(y)}
\nonumber\\&=&
D^0_{\mu\nu}(x,y)
+\int \dd z
\frac{\delta X_\mu(x)}{\delta \barA_\rho(z)}
\frac{\delta \barA_\rho(x)}{\delta \barj^\nu(y)}
\nonumber\\&=&
D^0_{\mu\nu}(x,y)
+\int \dd s \int \dd z
D^0_{\mu\sigma}(x,s)
\nonumber\\&&\times
\barPi^{\sigma\rho}(s,z)
\barD_{\rho\nu}(z,y),
\label{Dbar=D0Pi}
\end{eqnarray}
where we used 
\begin{eqnarray*}
\frac{\delta X_\mu(x)}{\delta \barA_\rho(z)}
&=& \int \dd s
D^0_{\mu\sigma}(x,s) \barPi^{\sigma\rho}(s,z).
\end{eqnarray*}
This last equation is proved by using
Eq.(\ref{Pi}) to rewrite
$X_\mu(x)$ in terms of a functional
derivative of $\barS$, and
by using Eq.(\ref{appendicederive}) 
to carry out the integral over $\lambda$.

Equation (\ref{Dbar=D0Pi}) expresses the usual
relation between the photon Green function and
the vacuum polarization.

\section{Weak field renormalization}
In this section, we give the renormalization rules of 
vacuum QED, which are valid for weak field renormalization.
We repeat some of the rules given in \cite{BFI} for
completeness, and because we work now in the direct
space (and not in the Fourier space).
The first set of rules is equivalent to 
Zimmermann's forest formula for the removal
of subdivergences. The second set
of rules subtracts the superficial divergences.

\subsection{Removal of subdivergences}
If $\Gamma$ is a Feynman diagram,
and $U(\Gamma)$ is the regularized value of the diagram
as a function of external momenta and masses and regularization
parameters, then the value of the
diagram with subdivergences subtracted is $\barR(\Gamma)$
defined by \cite{Itzykson}, \cite{Collins}
\begin{eqnarray}
\barR(\Gamma) &=& U(\Gamma)  +
 \sum_{\{\gamma i,\gamma' i',\dots\}}
C_i(\gamma)C_{i'}(\gamma')\cdots
  \nonumber\\&&\times
  U(\Gamma/\{\gamma_{(i)},\gamma'_{(i')},\dots\}).
\label{foret}
\end{eqnarray}

In this expression, the sum runs over all sets of disjoint
renormalization parts of $\Gamma$. A renormalization part $\gamma$ of
$\Gamma$ is a one-particle irreducible (1PI) subgraph of 
$\Gamma$, different from $\Gamma$ itself,
such that $\gamma$ has two or three (amputated) external lines.
A diagram is 1PI when it is connected and cannot be
disconnected by cutting through any of its internal lines.
Two renormalization parts $\gamma$ and $\gamma'$ are disjoint if
they have no vertex in common. 

There are three types of renormalization part:
self-energy 1PI diagrams
(i.e. 1PI diagrams with two amputed external electron lines),
vacuum polarization 1PI diagrams 
(i.e. 1PI diagrams with two amputed external photon lines)
and reduced vertex 1PI diagrams 
(i.e. 1PI diagrams with two amputed external electron lines
and one amputed external photon line).
The index $(i)$ depends on the type of the renormalization part.
If $\gamma$ is a self energy 1PI diagram then $i$ is 0 or 2,
if $\gamma$ is a vacuum polarization 1PI diagram then $i$ is 3,
if $\gamma$ is a reduced vertex 1PI diagram then $i$ is 1.

Finally let us define $\Gamma/\{\gamma_{(i)},\gamma'_{(i')},\dots\}$.
We start by the definition of $\Gamma/\gamma_{(i)}$.
It varies with the type of renormalization part and with the 
index $(i)$.

If $\gamma$ is a self-energy 1PI diagram, then
$i$ can be 0 or 2.
If the complete diagram is $\Gamma$ and $\gamma$
a self-energy 1PI subdiagram of $\Gamma$,
then in the term $\Gamma/\gamma_{(0)}$,
the 1PI diagram $\gamma$ is replaced by a point
and in the term
$\Gamma/\gamma_{(2)}$
the 1PI diagram $\gamma$ is replaced by a
free electron line.

If $\gamma$ is a vacuum polarization 1PI diagram, then
$i$ is 3.
If the complete diagram is $\Gamma$ and $\gamma$
is a vacuum polarization 1PI subdiagram of $\Gamma$,
then in the term $\Gamma/\gamma_{(3)}$,
the 1PI  diagram $\gamma$ is replaced
by a free photon propagator.

If $\gamma$ is a reduced vertex 1PI diagram, then
$i$ is 1.
If the complete diagram is $\Gamma$ and $\gamma$
is a reduced vertex 1PI diagram,
then in the term $\Gamma/\gamma_{(1)}$,
the vertex diagram is replaced
by a point.
The terms $\Gamma/\{\gamma_{(i)},\gamma'_{(i')},\dots\}$
are then defined recursively. For instance, to define
$\Gamma''=\Gamma/\{\gamma_{(i)},\gamma'_{(i')}\}$,
we first put
$\Gamma'=\Gamma/\gamma_{(i)}$, so that
$\Gamma''=\Gamma'/\gamma'_{(i')}$.

We insist here on the fact that the removal
of subdivergences given by Eq.(\ref{foret})
is valid for any (connected or disconnected)
Feynman diagram. It is the expression, for each
diagram, of Dyson's relations. The reason why
the counterterms $C_1(\gamma)$ are required, although
$Z_2=Z_1$, is explained in \cite{BFI}.

In Eq.(\ref{foret}) all counterterms 
$C_i(\gamma)$ are assumed to be known from the 
renormalization
of the superficial divergence of the subdiagram $\gamma$.

\subsection{Superficial divergences}
In the second step, we determine the counterterms $C_i(\Gamma)$
of the divergent graph $\Gamma$. 
If the diagram $\Gamma$ is not 1PI, the counterterms
$C_i(\Gamma)$ are zero.
If $\Gamma$ is 1PI, we must distinguish three cases.
A self-energy diagram is linearly divergent, thus we must
remove two terms. From Lorentz covariance, we can write the
renormalized value of the diagram $\Gamma$ as
\begin{eqnarray}
R(\Gamma;x,y) &=& \barR(\Gamma;x,y) + C_0(\Gamma) c^2 \delta(x-y)
\nonumber\\&& + C_2(\Gamma) {S^0}^{-1}(x,y).
\label{contretermese}
\end{eqnarray}

A vacuum polarization diagram is quadratically divergent, thus we
should have to remove three terms. However, Lorentz covariance
and the Ward identities for the photon propagators cancel the
first two counterterms, and the
renormalized value of $\Gamma$ is
\begin{eqnarray}
R_{\mu\nu}(\Gamma;x,y) &=& \barR_{\mu\nu}(\Gamma;x,y) 
+ C_3(\Gamma) {D^0}^{-1}_{\mu\nu}(x,y).
\label{contretermeva}
\end{eqnarray}
Finally, a reduced vertex diagram is logarithmically divergent,
and its renormalized value is
\begin{eqnarray}
R(\Gamma;x,y;\lambda,z) &=& \barR(\Gamma;x,y;\lambda,z) 
\nonumber\\&&\hspace*{-4mm}+ C_1(\Gamma)\gamma^\lambda 
\delta(x-y)\delta(y-z).
\label{contretermerv}
\end{eqnarray}
The infinite constants $C_i(\Gamma)$ are determined from the
renormalization conditions.

It was noticed in Ref.\cite{BFI} that
\begin{eqnarray}
C_2(\Gamma) &=& -\sum_j C_1(\Gamma_j), \label{C2C1}
\end{eqnarray}
where $\Gamma_j$ is the vertex diagram obtained 
by branching a photon
line on the $j$-th free electron propagator of 
the self-energy diagram $\Gamma$.

Since the renormalization of weak field QED is exactly the
same as the renormalization of vacuum QED, we refer the
reader to Refs.\cite{Itzykson} and \cite{BFI}
for examples.

\section{Strong field renormalization}
For the strong field renormalization, 
the removal of subdivergences is again given
by Eq.(\ref{foret}), which is a purely
algebraic identity related to the Hopf algebra
structure of renormalization \cite{CKI}, \cite{CKII}.

According to the general strategy of renormalization,
the non-renormalized diagrams are evaluated in terms
of renormalized electron propagators. In vacuum
QED, the mass $m$ used in the free electron
propagator is finite, it is not the infinite
mass $m_0$ of the bare free electron propagator.
Similarly, the charge and the potential used
in the basic electron propagator of strong
field QED are finite. In other words we
take
\begin{eqnarray*}
S^\bara &=& \big(i\hbar c \gamma\cdot\partial -mc^2-
  e\gamma\cdot \bara(x)\big)^{-1}.
\end{eqnarray*}
Therefore, the external potential diagram means now
\begin{eqnarray*}
U\big(
\parbox{5mm}{\begin{center}
\begin{fmfgraph}(5,2)
\setval
\fmfforce{0w,0.5h}{v1}
\fmfforce{1w,0.5h}{v2}
\fmf{boson}{v1,v2}
\fmfdot{v1}
\fmfv{decor.shape=pentagram,decor.filled=1,
      decor.size=2thick}{v2}
\end{fmfgraph}
\end{center}}\big) &=& \bara_\mu(x).
\end{eqnarray*}

To remove the superficial divergences
we start by
examining all possible superficially divergent diagrams
(Chapter 8 of Ref.\cite{Itzykson}).
A priori, the superficially divergent 1PI diagrams
are the self-energy and vertex diagrams,
and the 1PI diagrams with one external photon
(tadpole), two external photon (vacuum polarization),
three external photons and four external photons
(scattering of light by light).

\subsection{Self-energy and vertex diagrams}
To renormalize self-energy diagrams,
we take the example of the one-loop diagram, and we expand
the electron propagator using the Born expansion
(\cite{Schweber2} p.572)
\begin{eqnarray*}
S^\bara &=& S^0 + eS^0 \bara\cdot\gamma\, S^0 + 
e^2 S^0 \bara\cdot\gamma\, S^0
\bara\cdot\gamma\, S^\bara.
\end{eqnarray*}
Thus
\begin{eqnarray*}
\parbox{17mm}{\begin{center}
  \begin{fmfgraph}(15,5)
  \setval
  \fmfforce{0.2w,0.5h}{v1}
  \fmfforce{0.8w,0.5h}{v4}
  \fmf{dbl_plain}{v4,v1}
  \fmfdot{v1,v4}
  \fmf{boson,left}{v1,v4}
  \end{fmfgraph}
  \end{center}}
&=&
\parbox{17mm}{\begin{center}
  \begin{fmfgraph}(15,5)
  \setval
  \fmfforce{0.2w,0.5h}{v1}
  \fmfforce{0.8w,0.5h}{v4}
  \fmf{plain}{v4,v1}
  \fmf{boson,left}{v1,v4}
  \fmfdot{v1,v4}
  \end{fmfgraph}
  \end{center}}
+
e \parbox{17mm}{\begin{center}
  \begin{fmfgraph}(15,5)
  \setval
  \fmfforce{0.2w,0.5h}{v1}
  \fmfforce{0.5w,0.5h}{v2}
  \fmfforce{0.5w,0.0h}{vx}
  \fmfforce{0.8w,0.5h}{v4}
  \fmf{plain}{v4,v2,v1}
  \fmf{boson}{vx,v2}
  \fmf{boson,left}{v1,v4}
  \fmfdot{v1,v2,v4}
  \fmfv{decor.shape=pentagram,decor.filled=1,
      decor.size=2thick}{vx}
  \end{fmfgraph}
  \end{center}}
+ e^2 
\parbox{17mm}{\begin{center}
  \begin{fmfgraph}(15,5)
  \setval
  \fmfforce{0.0w,0.5h}{v1}
  \fmfforce{1/3w,0.5h}{v2}
  \fmfforce{1/3w,0.0h}{vx}
  \fmfforce{2/3w,0.5h}{v3}
  \fmfforce{2/3w,0.0h}{vy}
  \fmfforce{1w,0.5h}{v4}
  \fmf{plain}{v3,v2,v1}
  \fmf{dbl_plain}{v4,v3}
  \fmf{boson}{vx,v2}
  \fmf{boson}{vy,v3}
  \fmf{boson,left=0.6}{v1,v4}
  \fmfdot{v1,v2,v3,v4}
  \fmfv{decor.shape=pentagram,decor.filled=1,
      decor.size=2thick}{vx}
  \fmfv{decor.shape=pentagram,decor.filled=1,
      decor.size=2thick}{vy}
  \end{fmfgraph}
  \end{center}}.
\end{eqnarray*}
The strong field diagram on the left hand side
is denoted $\Gamma$.
The first diagram on the right hand side (denoted
$\Gamma'$) is a self-energy diagram of vacuum QED,
the second diagram (denoted $\Gamma'_1$) is a
vertex diagram of vacuum QED and the third diagram
is finite by power counting.
Using Eq.(\ref{contretermese}),
the counterterms for $\Gamma'$ are
\begin{eqnarray*}
R(\Gamma';x,y) &=& \barR(\Gamma';x,y) + C_0(\Gamma') c^2\delta(x-y)
\\&& + C_2(\Gamma') {S^0}^{-1}(x,y),
\end{eqnarray*}
Using Eq.(\ref{contretermerv}), the counterterm for the
second diagram is
\begin{eqnarray*}
\int \dd z R(\Gamma'_1;x,y;\lambda,z)\bara_\lambda(z) &=& 
\int \dd z \barR(\Gamma'_1;x,y;\lambda,z) \bara_\lambda(z)
\\&&+C_1(\Gamma'_1)\gamma^\lambda \delta(x-y)\bara_\lambda(y).
\end{eqnarray*}
The total counterterms for the strong field diagram $\Gamma$
are obtained by adding the counterterms for
$\Gamma'$ and $\Gamma'_1$.
From the relation (\ref{C2C1}) between 
$C_1(\Gamma'_1)$ and $C_2(\Gamma')$
we obtain for the strong field self-energy diagrams
\begin{eqnarray}
R(\Gamma;x,y) &=& \barR(\Gamma;x,y) + C_0(\Gamma) c^2 \delta(x-y)
\nonumber\\&&
 + C_2(\Gamma) ({S^0}^{-1}(x,y)-e \gamma\cdot \bara(y)\delta(x-y))
\nonumber\\
&=&
\barR(\Gamma;x,y) + C_0(\Gamma) c^2 \delta(x-y)
\nonumber\\&&
 + C_2(\Gamma) {S^\bara}^{-1}(x,y),
\label{SEfort}
\end{eqnarray}
where $C_0(\Gamma)=C_0(\Gamma')$
and $C_2(\Gamma)=C_2(\Gamma')$.
Therefore, the strong field self-energy diagrams are
renormalized with the same formula as the vacuum diagrams,
the counterterms $C_0(\Gamma)$ and $C_1(\Gamma)$ are the 
same as for the vacuum case, the only difference is that
the free propagator $S^0$ is replaced by the
strong field propagator $S^a$. More precisely, the
counterterms are the same as those of vacuum QED if
the renormalization conditions are compatible.
Otherwise, their difference is finite.

For a general strong field self-energy diagram $\Gamma$, 
the proof is similar. The subdivergences are removed
with formula (\ref{foret}). The same Born expansion is made.
The first diagram $\Gamma'$ is now the diagram
$\Gamma$ (with subtracted subdivergences) where all
strong field electron propagators
$S^a$ are replaced by free electron propagators $S^0$,
and $\Gamma'_j$ is the diagram obtained by adding 
a photon line to the $j$-th electron line of $\Gamma'$.
The remaining diagrams are finite by power counting.
The superficial divergences of $\Gamma'$ and
$\Gamma'_j$ are removed by the vacuum QED renormalization
prescription, and we find Eq.(\ref{SEfort}) again.

For a strong field vertex diagram, we have
a logarithmic divergence and the result is
exactly the same as for vacuum QED:
\begin{eqnarray}
R(\Gamma;x,y;\lambda,z) &=& \barR(\Gamma;x,y;\lambda,z) 
\nonumber\\&&+C_1(\Gamma)\gamma^\lambda \delta(x-y)\delta(y-z),
\label{vertexfort}
\end{eqnarray}
where $C_1(\Gamma)$ is the same as for vacuum QED,
up to possible finite terms.

\subsection{Vacuum polarization and tadpoles}

We still have the subdivergence-subtracted 1PI
diagrams with one, two, three or four
external photon lines (and no external electron line).
If the diagram has three or four external photon lines, then
it is finite. To prove this, we expand
the strong field electron propagator using the Born
expansion. Then we are back to the case of vacuum QED,
where we know that a 1PI diagram with three external photon lines
is zero and a 1PI with four external photon lines is
finite \cite{Itzykson} (once subdivergences are subracted).
A 1PI diagram with two external photon lines is a
vacuum polarization diagram, which is renormalized
as for vacuum QED with Eq.(\ref{contretermeva})
\begin{eqnarray}
R_{\mu\nu}(\Gamma;x,y) &=& \barR_{\mu\nu}(\Gamma;x,y) 
+ C_3(\Gamma) {D^0}^{-1}_{\mu\nu}(x,y),
\label{VPfort}
\end{eqnarray}
where $C_3(\Gamma)$ has the same value as for the
corresponding vacuum QED diagram, up to a possible
finite quantity.

Now comes the diagram which is really typical of
QED with external field: the tadpole.
The induced vector potential $A^\mu(x)-\bara^\mu(x)$
is given by the sum of all tadpole diagrams
$\Gamma$. In vacuum QED, the induced vector potential
is zero and the tadpoles do not intervene.
For the case of QED with an external field,
once subdivergences are removed,
the tadpole has a cubic superficial divergence.
But because of Furry's theorem (i.e. of the
fact that the current is odd under charge conjugation),
the only divergence that remains is the same as
for vacuum polarization.
We illustrate the general procedure with the simplest
example (see also \cite{Mohr}). By using the Born expansion 
for $S^\bara$ we obtain
\begin{eqnarray*}
\parbox{25mm}{\begin{center}
\begin{fmfgraph}(25,8)
\setval
\fmfforce{0.2w,0.5h}{v2}
\fmfforce{0.4w,8/16h}{v2b}
\fmfforce{0.6w,8/16h}{v3b}
\fmf{boson}{v2,v2b}
\fmf{dbl_plain,right}{v2b,v3b,v2b}
\fmfdot{v2,v2b}
\end{fmfgraph}
\end{center}}
&=&
\parbox{25mm}{\begin{center}
\begin{fmfgraph}(25,8)
\setval
\fmfforce{0.2w,0.5h}{v2}
\fmfforce{0.4w,8/16h}{v2b}
\fmfforce{0.6w,8/16h}{v3b}
\fmf{boson}{v2,v2b}
\fmf{plain,right}{v2b,v3b,v2b}
\fmfdot{v2,v2b}
\end{fmfgraph}
\end{center}}
+e
\parbox{25mm}{\begin{center}
\begin{fmfgraph}(25,8)
\setval
\fmfforce{0.2w,0.5h}{v2}
\fmfforce{0.4w,8/16h}{v2b}
\fmfforce{0.6w,8/16h}{v3b}
\fmfforce{0.8w,0.5h}{vx}
\fmf{boson}{v2,v2b}
\fmf{boson}{v3b,vx}
\fmf{plain,right}{v2b,v3b,v2b}
\fmfdot{v2,v2b,v3b}
\fmfv{decor.shape=pentagram,decor.filled=1,
      decor.size=2thick}{vx}
\end{fmfgraph}
\end{center}}
\\&&\hspace*{-8mm}
+ e^2
\parbox{25mm}{\begin{center}
\begin{fmfgraph}(25,8)
\setval
\fmfforce{0.2w,0.5h}{v2}
\fmfforce{0.4w,8/16h}{v2b}
\fmfforce{0.5w,3/16h}{vmb}
\fmfforce{0.5w,13/16h}{vmh}
\fmfforce{0.6w,0h}{vmbb}
\fmfforce{0.6w,1h}{vmhh}
\fmfforce{0.6w,8/16h}{v3b}
\fmf{boson}{v2,v2b}
\fmf{boson}{vmb,vmbb}
\fmf{boson}{vmh,vmhh}
\fmf{plain,right}{v2b,v3b,v2b}
\fmfdot{v2,v2b,vmb,vmh}
\fmfv{decor.shape=pentagram,decor.filled=1,
      decor.size=2thick}{vmbb,vmhh}
\end{fmfgraph}
\end{center}}
+ e^3
\parbox{25mm}{\begin{center}
\begin{fmfgraph}(25,8)
\setval
\fmfforce{0.2w,0.5h}{v2}
\fmfforce{0.4w,8/16h}{v2b}
\fmfforce{0.5w,3/16h}{vmb}
\fmfforce{0.5w,13/16h}{vmh}
\fmfforce{0.6w,0h}{vmbb}
\fmfforce{0.6w,1h}{vmhh}
\fmfforce{0.6w,8/16h}{v3b}
\fmfforce{0.8w,0.5h}{vx}
\fmf{boson}{v2,v2b}
\fmf{boson}{vmb,vmbb}
\fmf{boson}{vmh,vmhh}
\fmf{boson}{vx,v3b}
\fmf{plain}{v2b,vmh,v3b,vmb}
\fmf{dbl_plain}{vmb,v2b}
\fmfdot{v2,v2b,vmb,vmh,v3b}
\fmfv{decor.shape=pentagram,decor.filled=1,
      decor.size=2thick}{vmbb,vmhh,vx}
\end{fmfgraph}
\end{center}}
\,\,.
\end{eqnarray*}
Furry's theorem eliminates electron loops with one and
three external photon lines (the first and the third
diagrams on the right hand side). More precisely,
since the induced current is an odd function of
the external potential, the first and the third
diagrams are not present on the right hand side.
The last diagram
is finite, because if we make a further Born expansion
of $S^\bara$ in it, we obtain a vacuum photon-photon scattering
diagram, which is finite, plus a diagram with
five external photon lines, which is finite by power
counting. Therefore, the only divergent diagram on the rhs
is the second one, which is a vacuum polarization diagram.
We denote $\Gamma'_1$ this vacuum polarization diagram
and $\Gamma$ the strong field tadpole on the lhs.

We can write the bare potential as a sum of Feynman diagrams
\begin{eqnarray*}
A_\lambda(x) &=& a_\lambda(x) + \sum_{\Gamma} e_0^{|\Gamma|}
U_\lambda(x;\Gamma),
\end{eqnarray*}
where $|\Gamma|$ is the number of vertices of the
tadpole diagram $\Gamma$.
If $\Gamma$ is the simple tadpole of the above example,
our discussion shows that its value is
\begin{eqnarray*}
U_\lambda(x;\Gamma) &=&
e\int \dd y\dd z D^0_{\lambda\mu}(x,y) 
\barR^{\mu\nu}(\Gamma'_1;y,z) \bara_\nu(z)
\\&&
+\mathrm{finite}\,\,\mathrm{terms}. 
\end{eqnarray*}

Now, we introduce the counterterm of
$\Gamma'_1$ using Eq.(\ref{contretermeva}),
and we obtain the renormalized
induced potential
\begin{eqnarray*}
R_\lambda(x;\Gamma) &=&
e\int \dd y\dd z D^0_{\lambda\mu}(x,y)
\barR^{\mu\nu}(\Gamma'_1;y,z) \bara_\nu(z)
\\&&
+eC_3(\Gamma'_1)\bara_\lambda(x)
+\mathrm{finite}\,\,\mathrm{terms}.
\end{eqnarray*}
Similar equations can be found
in \cite{Schweber2} p. 552, \cite{Engel4}.

Therefore, the counterterm for the tadpole diagram
$\Gamma$ is
\begin{eqnarray*}
R_\lambda(\Gamma;x) &=& \barR_\lambda(\Gamma;x) + e C_4(\Gamma)\bara_\lambda(x),
\end{eqnarray*}
where $C_4(\Gamma)=C_3(\Gamma'_1)$.

For a general tadpole diagram $\Gamma$, the result is
similar. If $\barR^\mu(\Gamma;x)$ is the value of 
$\Gamma$ when all subdivergences are removed, the 
superficial divergence is removed with the counterterm
\begin{eqnarray}
R^\mu(\Gamma;x) &=& \barR^\mu(\Gamma;x) + e C_4(\Gamma)\bara^\mu(x),
\label{rentadpole}
\end{eqnarray}
where
\begin{eqnarray*}
C_4(\Gamma) &=& \sum_i C_3(\Gamma_i).
\end{eqnarray*}
In the last relation, $\Gamma_i$ is the vacuum
polarization obtained from the tadpole
$\Gamma$ by adding a photon line to the $i$-th
electron line of $\Gamma$ and 
by transforming all electron lines
$S^a$ into free electron lines
$S^0$ in the resulting diagram. 
In other words, $\Gamma_i$ is a vacuum QED
vacuum polarization diagram, and $C_3(\Gamma_i)$
the corresponding counterterm.

After this analysis, we obtain the following 
addition to the description of the renormalization
of vacuum QED. There is now a new index $i=4$
and new 1PI diagrams $\gamma$ which are tadpoles.
In the term $\Gamma/\gamma_{(4)}$, the tadpole
subdiagram $\gamma_{(4)}$ becomes an
external potential line
$\parbox{5mm}{\begin{center}
\begin{fmfgraph}(5,2)
\setval
\fmfforce{0w,0.5h}{v1}
\fmfforce{1w,0.5h}{v2}
\fmf{boson}{v1,v2}
\fmfdot{v1}
\fmfv{decor.shape=pentagram,decor.filled=1,
      decor.size=2thick}{v2}
\end{fmfgraph}
\end{center}}$\,.

\subsection{Summary}
In this section, we summarize the renormalization
rules for strong field QED.
For any strong field Feynman diagram $\Gamma$,
the subdivergences are subtracted by applying
the forest formula (\ref{foret}).
The only difference with the weak field case
is that the vacuum electron propagators are
replaced by strong field electron propagators.

When the subdivergences are subtracted,
the superficial divergence is treated 
as follows. If $\Gamma$ is not 1PI, the
superficial counterterms are zero. In particular,
if $\Gamma$ contains tadpoles,
then the counterterms $C_i(\Gamma)$ are zero.
This is because a tadpole inside $\Gamma$ can always be 
disconnected from $\Gamma$ by cutting a photon line.

If $\Gamma$ is 1PI, then the superficial divergence
is removed using Eq.(\ref{SEfort}) for a self-energy and
the indices are $i=0$ and $i=2$,
Eq.(\ref{vertexfort}) for a vertex and the index is $i=1$,
Eq.(\ref{VPfort}) for a vacuum polarization with the
index $i=3$
and Eq.(\ref{rentadpole}) for a tadpole with the index $i=4$.

Now we give a few example to illustrate the general rules.

\subsection{Example 1: the potential}
As far as we know, the renormalization rules for
strong field QED are new. Therefore, a
number of examples are required to see
how they work in practice.

We start with the renormalization of the
potential for the diagrams given in
section \ref{FeynStrong}.
The first diagram has no subdivergence.
It is renormalized with Eq.(\ref{rentadpole}).
\begin{eqnarray*}
R\big(
\parbox{11mm}{\begin{center}
\begin{fmfgraph}(10,5)
\setval
\fmfforce{0w,0.5h}{v1}
\fmfforce{1/2w,0.5h}{v2}
\fmfforce{1w,0.5h}{v3}
\fmf{dbl_plain,right}{v2,v3,v2}
\fmf{boson}{v1,v2}
\fmfdot{v1,v2}
\end{fmfgraph}
\end{center}} \big)
&=&
U\big(
\parbox{11mm}{\begin{center}
\begin{fmfgraph}(10,5)
\setval
\fmfforce{0w,0.5h}{v1}
\fmfforce{1/2w,0.5h}{v2}
\fmfforce{1w,0.5h}{v3}
\fmf{dbl_plain,right}{v2,v3,v2}
\fmf{boson}{v1,v2}
\fmfdot{v1,v2}
\end{fmfgraph}
\end{center}} \big)
+e C_4\big(
\parbox{11mm}{\begin{center}
\begin{fmfgraph}(10,5)
\setval
\fmfforce{0w,0.5h}{v1}
\fmfforce{1/2w,0.5h}{v2}
\fmfforce{1w,0.5h}{v3}
\fmf{dbl_plain,right}{v2,v3,v2}
\fmf{boson}{v1,v2}
\fmfdot{v1,v2}
\end{fmfgraph}
\end{center}} \big)
U\big(
\parbox{5mm}{\begin{center}
\begin{fmfgraph}(5,2)
\setval
\fmfforce{0w,0.5h}{v1}
\fmfforce{1w,0.5h}{v2}
\fmf{boson}{v1,v2}
\fmfdot{v1}
\fmfv{decor.shape=pentagram,decor.filled=1,
      decor.size=2thick}{v2}
\end{fmfgraph}
\end{center}}\big).
\end{eqnarray*}

The second diagram is not 1PI. To renormalize it,
we renormalize its two disjoint divergent subdiagrams:
the vacuum polarization diagram
on the left (index $i=3$) and the tadpole on the 
right (index $i=4$), and we
multiply the renormalized diagrams.
The general rule for a non 1PI diagram $\Gamma$
is the following: write $\Gamma$ as a product
of 1PI diagrams, renormalized each 1PI diagram,
and multiply the renormalized diagrams \cite{Bogoliubov}.
This rule is equivalent to the forest
formula (\ref{foret}) for non 1PI diagrams,
because renormalization parts belonging to
different 1PI subdiagrams are disjoint.

\begin{eqnarray*}
R\big(
\parbox{21mm}{\begin{center}
\begin{fmfgraph}(20,5)
\setval
\fmfforce{0w,0.5h}{v2}
\fmfforce{1/4w,0.5h}{v3}
\fmfforce{2/4w,0.5h}{v4}
\fmfforce{3/4w,0.5h}{v5}
\fmfforce{1w,0.5h}{v6}
\fmf{dbl_plain,right}{v3,v4,v3}
\fmf{boson}{v2,v3}
\fmf{dbl_plain,right}{v5,v6,v5}
\fmf{boson}{v4,v5}
\fmfdot{v2,v3,v4,v5}
\end{fmfgraph}
\end{center}} \big)
&=&
U\big(
\parbox{21mm}{\begin{center}
\begin{fmfgraph}(20,5)
\setval
\fmfforce{0w,0.5h}{v2}
\fmfforce{1/4w,0.5h}{v3}
\fmfforce{2/4w,0.5h}{v4}
\fmfforce{3/4w,0.5h}{v5}
\fmfforce{1w,0.5h}{v6}
\fmf{dbl_plain,right}{v3,v4,v3}
\fmf{boson}{v2,v3}
\fmf{dbl_plain,right}{v5,v6,v5}
\fmf{boson}{v4,v5}
\fmfdot{v2,v3,v4,v5}
\end{fmfgraph}
\end{center}} \big)
\\&&\hspace*{-25mm}
+C_3\big(
\parbox{7mm}{\begin{center}
\begin{fmfgraph}(6,4)
\setval
\fmfforce{0w,0.5h}{v1}
\fmfforce{4/24w,0.5h}{v2}
\fmfforce{20/24w,0.5h}{v3}
\fmfforce{1w,0.5h}{v4}
\fmf{boson}{v1,v2}
\fmf{dbl_plain,left}{v2,v3,v2}
\fmf{boson}{v3,v4}
\fmfdot{v2,v3}
\end{fmfgraph}
\end{center}}
\big)
U\big(
\parbox{11mm}{\begin{center}
\begin{fmfgraph}(10,5)
\setval
\fmfforce{0w,0.5h}{v1}
\fmfforce{1/2w,0.5h}{v2}
\fmfforce{1w,0.5h}{v3}
\fmf{dbl_plain,right}{v2,v3,v2}
\fmf{boson}{v1,v2}
\fmfdot{v1,v2}
\end{fmfgraph}
\end{center}} \big)
+e C_4\big(
\parbox{11mm}{\begin{center}
\begin{fmfgraph}(10,5)
\setval
\fmfforce{0w,0.5h}{v1}
\fmfforce{1/2w,0.5h}{v2}
\fmfforce{1w,0.5h}{v3}
\fmf{dbl_plain,right}{v2,v3,v2}
\fmf{boson}{v1,v2}
\fmfdot{v1,v2}
\end{fmfgraph}
\end{center}} \big)
U\big(
\parbox{11mm}{\begin{center}
\begin{fmfgraph}(10,5)
\setval
\fmfforce{0w,0.5h}{v1}
\fmfforce{1/4w,0.5h}{v2}
\fmfforce{3/4w,0.5h}{v3}
\fmfforce{1w,0.5h}{v4}
\fmf{dbl_plain,right}{v2,v3,v2}
\fmf{boson}{v1,v2}
\fmf{boson}{v3,v4}
\fmfdot{v1,v2,v3}
\fmfv{decor.shape=pentagram,decor.filled=1,
      decor.size=2thick}{v4}
\end{fmfgraph}
\end{center}} \big)
\\&&\hspace*{-25mm}
+
e C_3\big(
\parbox{7mm}{\begin{center}
\begin{fmfgraph}(6,4)
\setval
\fmfforce{0w,0.5h}{v1}
\fmfforce{4/24w,0.5h}{v2}
\fmfforce{20/24w,0.5h}{v3}
\fmfforce{1w,0.5h}{v4}
\fmf{boson}{v1,v2}
\fmf{dbl_plain,left}{v2,v3,v2}
\fmf{boson}{v3,v4}
\fmfdot{v2,v3}
\end{fmfgraph}
\end{center}}
\big)
C_4\big(
\parbox{11mm}{\begin{center}
\begin{fmfgraph}(10,5)
\setval
\fmfforce{0w,0.5h}{v1}
\fmfforce{1/2w,0.5h}{v2}
\fmfforce{1w,0.5h}{v3}
\fmf{dbl_plain,right}{v2,v3,v2}
\fmf{boson}{v1,v2}
\fmfdot{v1,v2}
\end{fmfgraph}
\end{center}} \big)
U\big(
\parbox{5mm}{\begin{center}
\begin{fmfgraph}(5,2)
\setval
\fmfforce{0w,0.5h}{v1}
\fmfforce{1w,0.5h}{v2}
\fmf{boson}{v1,v2}
\fmfdot{v1}
\fmfv{decor.shape=pentagram,decor.filled=1,
      decor.size=2thick}{v2}
\end{fmfgraph}
\end{center}}\big).
\end{eqnarray*}

For the last diagram, there are two subdivergences:
on the left, a self-energy $\gamma$ (indices $i=0$ and $i=2$)
and on the right, a vertex $\gamma'$ (index $i=1$).
Since these subdiagrams are not disjoint we obtain,
adding the superficial divergence counterterm
\begin{eqnarray*}
R\big(
\parbox{11mm}{\begin{center}
\begin{fmfgraph}(10,5)
\setval
\fmfforce{0w,0.5h}{v1}
\fmfforce{1/2w,0.5h}{v2}
\fmfforce{3/4w,1h}{vx}
\fmfforce{3/4w,0h}{vy}
\fmfforce{1w,0.5h}{v3}
\fmf{dbl_plain,right}{v2,v3,v2}
\fmf{boson}{v1,v2}
\fmf{boson}{vx,vy}
\fmfdot{v1,v2,vx,vy}
\end{fmfgraph}
\end{center}} \big)
&=&
U\big(
\parbox{11mm}{\begin{center}
\begin{fmfgraph}(10,5)
\setval
\fmfforce{0w,0.5h}{v1}
\fmfforce{1/2w,0.5h}{v2}
\fmfforce{3/4w,1h}{vx}
\fmfforce{3/4w,0h}{vy}
\fmfforce{1w,0.5h}{v3}
\fmf{dbl_plain,right}{v2,v3,v2}
\fmf{boson}{v1,v2}
\fmf{boson}{vx,vy}
\fmfdot{v1,v2,vx,vy}
\end{fmfgraph}
\end{center}} \big)
+e C_4\big(
\parbox{11mm}{\begin{center}
\begin{fmfgraph}(10,5)
\setval
\fmfforce{0w,0.5h}{v1}
\fmfforce{1/2w,0.5h}{v2}
\fmfforce{3/4w,1h}{vx}
\fmfforce{3/4w,0h}{vy}
\fmfforce{1w,0.5h}{v3}
\fmf{dbl_plain,right}{v2,v3,v2}
\fmf{boson}{v1,v2}
\fmf{boson}{vx,vy}
\fmfdot{v1,v2,vx,vy}
\end{fmfgraph}
\end{center}} \big)
U\big(
\parbox{5mm}{\begin{center}
\begin{fmfgraph}(5,2)
\setval
\fmfforce{0w,0.5h}{v1}
\fmfforce{1w,0.5h}{v2}
\fmf{boson}{v1,v2}
\fmfdot{v1}
\fmfv{decor.shape=pentagram,decor.filled=1,
      decor.size=2thick}{v2}
\end{fmfgraph}
\end{center}}
\big)
\\&&\hspace*{-5mm}
+C_0\big(
\parbox{5mm}{\begin{center}\begin{fmfgraph}(4,3)
\setval
\fmfforce{0w,0.5h}{v1}
\fmfforce{1w,0.5h}{v2}
\fmf{dbl_plain}{v1,v2}
\fmf{boson,left=1}{v1,v2}
\fmfdot{v1,v2}
\end{fmfgraph}
\end{center}}\big)
U\big(
\parbox{11mm}{\begin{center}
\begin{fmfgraph}(10,5)
\setval
\fmfforce{0w,0.5h}{v1}
\fmfforce{1/2w,0.5h}{v2}
\fmfforce{1w,0.5h}{v3}
\fmf{dbl_plain,right}{v2,v3,v2}
\fmf{boson}{v1,v2}
\fmfdot{v1,v2,v3}
\end{fmfgraph}
\end{center}} \big)
+C_2\big(
\parbox{5mm}{\begin{center}\begin{fmfgraph}(4,3)
\setval
\fmfforce{0w,0.5h}{v1}
\fmfforce{1w,0.5h}{v2}
\fmf{dbl_plain}{v1,v2}
\fmf{boson,left=1}{v1,v2}
\fmfdot{v1,v2}
\end{fmfgraph}
\end{center}}\big)
U\big(
\parbox{11mm}{\begin{center}
\begin{fmfgraph}(10,5)
\setval
\fmfforce{0w,0.5h}{v1}
\fmfforce{1/2w,0.5h}{v2}
\fmfforce{1w,0.5h}{v3}
\fmf{dbl_plain,right}{v2,v3,v2}
\fmf{boson}{v1,v2}
\fmfdot{v1,v2}
\end{fmfgraph}
\end{center}} \big)
\\&&\hspace*{-5mm}
+C_1\big(
\parbox{5mm}{\begin{center}\begin{fmfgraph}(4,3)
\setval
\fmfforce{0w,0.5h}{v1}
\fmfforce{1w,0.5h}{v2}
\fmfforce{0.5w,0.5h}{vx}
\fmfforce{0.5w,0.1h}{vy}
\fmf{dbl_plain}{v1,vx,v2}
\fmf{boson,left=1}{v1,v2}
\fmf{boson}{vx,vy}
\fmfdot{v1,v2,vx}
\end{fmfgraph}
\end{center}}\big)
U\big(
\parbox{11mm}{\begin{center}
\begin{fmfgraph}(10,5)
\setval
\fmfforce{0w,0.5h}{v1}
\fmfforce{1/2w,0.5h}{v2}
\fmfforce{1w,0.5h}{v3}
\fmf{dbl_plain,right}{v2,v3,v2}
\fmf{boson}{v1,v2}
\fmfdot{v1,v2}
\end{fmfgraph}
\end{center}} \big).
\end{eqnarray*}

To check these results, we make use of the
renormalization factors.
\begin{eqnarray*}
Z_2 &=& 1+\sum_{n=1}^\infty e^{2n} Z_{2,n},\\
Z_3 &=& 1-\sum_{n=1}^\infty e^{2n} Z_{3,n},\\
\delta m &=& \sum_{n=1}^\infty e^{2n} \delta m_{n}.
\end{eqnarray*}
The relation between the renormalization factors
and the counterterms is as follows \cite{BFI}.
$Z_{2,n}$ is the sum of all $C_1(\Gamma)$ where
$\Gamma$ runs over the (1PI) vertex diagrams with 
$n$ loops, $Z_{2,n}$ is also the sum of all 
$-C_2(\Gamma)$ where
$\Gamma$ runs over the (1PI) self-energy diagrams with 
$n$ loops,
$Z_{3,n}$ is the sum of all 
$C_3(\Gamma)$ where
$\Gamma$ runs over the (1PI) vacuum polarization diagrams with
$n$ loops, $Z_{3,n}$ is also the sum of all      
$C_4(\Gamma)$ where
$\Gamma$ runs over the (1PI) tadpole diagrams with
$n$ loops. Finally, 
$\delta m_n + \sum_{k=1}^{n-1} \delta m_k Z_{2,n-k}$ is the
sum of all $C_0(\Gamma)$ where
$\Gamma$ runs over the (1PI) self-energy diagrams with
$n$ loops.

For our example, we have
\begin{eqnarray*}
Z_{2,1} &=&
C_1\big(
\parbox{5mm}{\begin{center}\begin{fmfgraph}(4,3)
\setval
\fmfforce{0w,0.5h}{v1}
\fmfforce{1w,0.5h}{v2}
\fmfforce{0.5w,0.5h}{vx}
\fmfforce{0.5w,0.1h}{vy}
\fmf{dbl_plain}{v1,vx,v2}
\fmf{boson,left=1}{v1,v2}
\fmf{boson}{vx,vy}
\fmfdot{v1,v2,vx}
\end{fmfgraph}
\end{center}}\big)
=
-C_2\big(
\parbox{5mm}{\begin{center}\begin{fmfgraph}(4,3)
\setval
\fmfforce{0w,0.5h}{v1}
\fmfforce{1w,0.5h}{v2}
\fmf{dbl_plain}{v1,v2}
\fmf{boson,left=1}{v1,v2}
\fmfdot{v1,v2}
\end{fmfgraph}
\end{center}}\big),
\\
Z_{3,1} &=&
C_3\big(
\parbox{7mm}{\begin{center}
\begin{fmfgraph}(6,4)
\setval
\fmfforce{0w,0.5h}{v1}
\fmfforce{4/24w,0.5h}{v2}
\fmfforce{20/24w,0.5h}{v3}
\fmfforce{1w,0.5h}{v4}
\fmf{boson}{v1,v2}
\fmf{dbl_plain,left}{v2,v3,v2}
\fmf{boson}{v3,v4}
\fmfdot{v2,v3}
\end{fmfgraph}
\end{center}}
\big)
=
C_4\big(
\parbox{9mm}{\begin{center}
\begin{fmfgraph}(8,4)
\setval
\fmfforce{0w,0.5h}{v1}
\fmfforce{1/2w,0.5h}{v2}
\fmfforce{1w,0.5h}{v3}
\fmf{dbl_plain,right}{v2,v3,v2}
\fmf{boson}{v1,v2}
\fmfdot{v1,v2}
\end{fmfgraph}
\end{center}} \big),\\
\delta m_1 &=&
C_0\big(
\parbox{5mm}{\begin{center}\begin{fmfgraph}(4,3)
\setval
\fmfforce{0w,0.5h}{v1}
\fmfforce{1w,0.5h}{v2}
\fmf{dbl_plain}{v1,v2}
\fmf{boson,left=1}{v1,v2}
\fmfdot{v1,v2}
\end{fmfgraph}
\end{center}}\big).
\end{eqnarray*}

Thus, the renormalized induced potential up to two loops is
\begin{eqnarray*}
e
R\big(
\parbox{11mm}{\begin{center}
\begin{fmfgraph}(10,5)
\setval
\fmfforce{0w,0.5h}{v1}
\fmfforce{1/2w,0.5h}{v2}
\fmfforce{1w,0.5h}{v3}
\fmf{dbl_plain,right}{v2,v3,v2}
\fmf{boson}{v1,v2}
\fmfdot{v1,v2}
\end{fmfgraph}
\end{center}} \big)
+
e^3 R\big(
\parbox{21mm}{\begin{center}
\begin{fmfgraph}(20,5)
\setval
\fmfforce{0w,0.5h}{v2}
\fmfforce{1/4w,0.5h}{v3}
\fmfforce{2/4w,0.5h}{v4}
\fmfforce{3/4w,0.5h}{v5}
\fmfforce{1w,0.5h}{v6}
\fmf{dbl_plain,right}{v3,v4,v3}
\fmf{boson}{v2,v3}
\fmf{dbl_plain,right}{v5,v6,v5}
\fmf{boson}{v4,v5}
\fmfdot{v2,v3,v4,v5}
\end{fmfgraph}
\end{center}} \big)
+
e^3 R\big(
\parbox{11mm}{\begin{center}
\begin{fmfgraph}(10,5)
\setval
\fmfforce{0w,0.5h}{v1}
\fmfforce{1/2w,0.5h}{v2}
\fmfforce{3/4w,1h}{vx}
\fmfforce{3/4w,0h}{vy}
\fmfforce{1w,0.5h}{v3}
\fmf{dbl_plain,right}{v2,v3,v2}
\fmf{boson}{v1,v2}
\fmf{boson}{vx,vy}
\fmfdot{v1,v2,vx,vy}
\end{fmfgraph}
\end{center}} \big)
&=&
\\&&\hspace*{-75mm}
e U\big(
\parbox{11mm}{\begin{center}
\begin{fmfgraph}(10,5)
\setval
\fmfforce{0w,0.5h}{v1}
\fmfforce{1/2w,0.5h}{v2}
\fmfforce{1w,0.5h}{v3}
\fmf{dbl_plain,right}{v2,v3,v2}
\fmf{boson}{v1,v2}
\fmfdot{v1,v2}
\end{fmfgraph}
\end{center}} \big)
+
e^3 U\big(
\parbox{21mm}{\begin{center}
\begin{fmfgraph}(20,5)
\setval
\fmfforce{0w,0.5h}{v2}
\fmfforce{1/4w,0.5h}{v3}
\fmfforce{2/4w,0.5h}{v4}
\fmfforce{3/4w,0.5h}{v5}
\fmfforce{1w,0.5h}{v6}
\fmf{dbl_plain,right}{v3,v4,v3}
\fmf{boson}{v2,v3}
\fmf{dbl_plain,right}{v5,v6,v5}
\fmf{boson}{v4,v5}
\fmfdot{v2,v3,v4,v5}
\end{fmfgraph}
\end{center}} \big)
+
e^3 U\big(
\parbox{11mm}{\begin{center}
\begin{fmfgraph}(10,5)
\setval
\fmfforce{0w,0.5h}{v1}
\fmfforce{1/2w,0.5h}{v2}
\fmfforce{3/4w,1h}{vx}
\fmfforce{3/4w,0h}{vy}
\fmfforce{1w,0.5h}{v3}
\fmf{dbl_plain,right}{v2,v3,v2}
\fmf{boson}{v1,v2}
\fmf{boson}{vx,vy}
\fmfdot{v1,v2,vx,vy}
\end{fmfgraph}
\end{center}} \big)
\\&&\hspace*{-75mm}
+
\big( e^2 Z_{3,1} + e^4 (Z_{3,2} + Z_{3,1}^2) \big)
U\big(
\parbox{5mm}{\begin{center}
\begin{fmfgraph}(5,2)
\setval
\fmfforce{0w,0.5h}{v1}
\fmfforce{1w,0.5h}{v2}
\fmf{boson}{v1,v2}
\fmfdot{v1}
\fmfv{decor.shape=pentagram,decor.filled=1,
      decor.size=2thick}{v2}
\end{fmfgraph}
\end{center}}
\big)
\\&&\hspace*{-75mm}
+e^3 Z_{3,1}
U\big(
\parbox{11mm}{\begin{center}
\begin{fmfgraph}(10,5)
\setval
\fmfforce{0w,0.5h}{v1}
\fmfforce{1/2w,0.5h}{v2}
\fmfforce{1w,0.5h}{v3}
\fmf{dbl_plain,right}{v2,v3,v2}
\fmf{boson}{v1,v2}
\fmfdot{v1,v2}
\end{fmfgraph}
\end{center}} \big)
+e^3 Z_{3,1}
U\big(
\parbox{11mm}{\begin{center}
\begin{fmfgraph}(10,5)
\setval
\fmfforce{0w,0.5h}{v1}
\fmfforce{1/4w,0.5h}{v2}
\fmfforce{3/4w,0.5h}{v3}
\fmfforce{1w,0.5h}{v4}
\fmf{dbl_plain,right}{v2,v3,v2}
\fmf{boson}{v1,v2}
\fmf{boson}{v3,v4}
\fmfdot{v1,v2,v3}
\fmfv{decor.shape=pentagram,decor.filled=1,
      decor.size=2thick}{v4}
\end{fmfgraph}
\end{center}} \big)
\\&&\hspace*{-75mm}
+e^3 \delta m_1
U\big(
\parbox{11mm}{\begin{center}
\begin{fmfgraph}(10,5)
\setval
\fmfforce{0w,0.5h}{v1}
\fmfforce{1/2w,0.5h}{v2}
\fmfforce{1w,0.5h}{v3}
\fmf{dbl_plain,right}{v2,v3,v2}
\fmf{boson}{v1,v2}
\fmfdot{v1,v2,v3}
\end{fmfgraph}
\end{center}} \big).
\end{eqnarray*}

To check this result, we take Dyson's
relation between renormalized and bare
electron Green functions
\begin{eqnarray*}
Z_2(e,m) \barS(x,y;e\bara;e,m)
&=&
S(x,y;\frac{e\bara}{Z_3};\frac{e}{\sqrt{Z_3}},m+\delta m).
\end{eqnarray*}
We expand the right hand side
\begin{eqnarray*}
Z_2(e,m) \barS(x,y;e\bara;e,m)
&=&
U\big(
\parbox{5mm}{\begin{center}\begin{fmfgraph}(4,3)
\setval
\fmfforce{0w,0.5h}{v1}
\fmfforce{1w,0.5h}{v2}
\fmf{dbl_plain}{v1,v2}
\fmfdot{v1,v2}
\end{fmfgraph}
\end{center}}\big)
+e^2 Z_{3,1}
U\big(
\parbox{6mm}{\begin{center}\begin{fmfgraph}(5,3)
\setval
\fmfforce{0w,0.5h}{v1}
\fmfforce{0.5w,0.5h}{vx}
\fmfforce{0.5w,0.0h}{vy}
\fmfforce{1w,0.5h}{v2}
\fmf{dbl_plain}{v1,vx,v2}
\fmf{boson}{vx,vy}
\fmfdot{v1,v2,vx}
\fmfv{decor.shape=pentagram,decor.filled=1,
      decor.size=2thick}{vy}
\end{fmfgraph}
\end{center}}\big)
\\&&\hspace*{-30mm}
+e^2 \delta m_1
U\big(
\parbox{6mm}{\begin{center}\begin{fmfgraph}(5,3)
\setval
\fmfforce{0w,0.5h}{v1}
\fmfforce{0.5w,0.5h}{vx}
\fmfforce{1w,0.5h}{v2}
\fmf{dbl_plain}{v1,vx,v2}
\fmfdot{v1,v2,vx}
\end{fmfgraph}
\end{center}}\big)
+e^2 U\big(
 \parbox{21mm}{\begin{center}
  \begin{fmfgraph}(20,5)
  \setval
  \fmfforce{0w,0.5h}{v1}
  \fmfforce{1/3w,0.5h}{v2}
  \fmfforce{2/3w,0.5h}{v3}
  \fmfforce{1w,0.5h}{v4}
  \fmf{dbl_plain}{v4,v3,v2,v1}
  \fmf{boson,left}{v2,v3}
  \fmfdot{v1,v2,v3,v4}
  \end{fmfgraph}
  \end{center}}\big)
\\&&\hspace*{-30mm}
+e^2 U\big(
\parbox{17mm}{\begin{center}
\begin{fmfgraph}(16,8)
\setval
\fmfforce{0w,0h}{v2}
\fmfforce{0.4w,11/16h}{v2b}
\fmfforce{0.5w,0h}{vx}
\fmfforce{0.5w,39/80h}{vxb}
\fmfforce{0.6w,11/16h}{v3b}
\fmfforce{1w,0h}{v3}
\fmf{dbl_plain}{v3,vx,v2}
\fmf{boson}{vx,vxb}
\fmf{dbl_plain,right}{v2b,v3b,v2b}
\fmfdot{v2,vx,vxb,v3}
\end{fmfgraph}
\end{center}}\big)
+\cdots
\end{eqnarray*}
where the electron propagator
$\parbox{5mm}{\begin{center}\begin{fmfgraph}(4,3)
\setval
\fmfforce{0w,0.5h}{v1}
\fmfforce{1w,0.5h}{v2}
\fmf{dbl_plain}{v1,v2}
\fmfdot{v1,v2}
\end{fmfgraph}
\end{center}}$ is calculated
in the field of the renormalized external
potentials $\bara(x)$.
If we use this expansion in Eq.(\ref{AZ2Z3})
we obtain the renormalized potential up to two loops.
We recall that
\begin{eqnarray*}
U\big(
\parbox{6mm}{\begin{center}\begin{fmfgraph}(5,3)
\setval
\fmfforce{0w,0.5h}{v1}
\fmfforce{0.5w,0.5h}{vx}
\fmfforce{1w,0.5h}{v2}
\fmf{dbl_plain}{v1,vx,v2}
\fmfdot{v1,v2,vx}
\end{fmfgraph}
\end{center}}\big)
&=&
\int \dd z S^a(x,z)S^a(z,y)
=\frac{\partial S^a(x,y)}{\partial m}.
\end{eqnarray*}

Obviously, it is much simpler to renormalize
directly from the Dyson relation. However, this does 
not renormalized all Feynman diagrams. Only the
sum of them is finite, as was discussed in
\cite{BFI}. In practice, all the Feynman diagrams
of a certain order cannot always be calculated, and
it is necessary to renormalize each diagram separately.

\subsection{Example 2: the self-energy}
For QED in an external field, the bare 
and renormalized self-energies are defined by
\begin{eqnarray*}
S^{-1} &=& i\hbar c\gamma\cdot\partial - m_0 -e_0\gamma\cdot A
-\Sigma,\\
\barS^{-1} &=& i\hbar c\gamma\cdot\partial - m -e\gamma\cdot \barA
-\bar\Sigma.
\end{eqnarray*}
From Dyson's relations between $S$ and $\barS$ we obtain
the following relation between $\Sigma$ and $\bar\Sigma$.
\begin{eqnarray*}
\bar\Sigma(e\bara;e,m) &=&
Z_2 \Sigma(e\bara/Z_3;e/\sqrt{Z_3},m+\delta m)
+Z_2\delta m 
\\&&+
(Z_2-1)(-i\hbar c\gamma\cdot\partial + m + e\gamma\cdot \barA).
\end{eqnarray*}

In the self-energy, the only new diagram with respect
to vacuum QED is
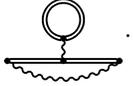
\begin{eqnarray*}
\parbox{16mm}{\begin{center}
\begin{fmfgraph}(15,8)
\setval
\fmfforce{0w,0h}{v2}
\fmfforce{0.5w,0h}{vx}
\fmfforce{0.5w,6/16h}{vxb}
\fmfforce{0.5w,1h}{vxc}
\fmfforce{1w,0h}{v3}
\fmf{dbl_plain}{v3,v2}
\fmf{boson,right=0.33}{v2,v3}
\fmf{boson}{vx,vxb}
\fmf{dbl_plain,right}{vxb,vxc,vxb}
\fmfdot{v2,v3,vx,vxb}
\end{fmfgraph}
\end{center}}.
\end{eqnarray*}
This diagram is not 1PI. In vacuum QED, all self-energy
diagrams are 1PI, but this is not the case in the
presence of an external field. This diagram is easily
renormalized as a product of two 1PI diagrams:
\begin{eqnarray*}
R\big(
\parbox{11mm}{\begin{center}
\begin{fmfgraph}(10,4)
\setval
\fmfforce{0w,0h}{v2}
\fmfforce{0.5w,0h}{vx}
\fmfforce{0.5w,6/16h}{vxb}
\fmfforce{0.5w,1h}{vxc}
\fmfforce{1w,0h}{v3}
\fmf{dbl_plain}{v3,v2}
\fmf{boson,right=0.33}{v2,v3}
\fmf{boson}{vx,vxb}
\fmf{dbl_plain,right}{vxb,vxc,vxb}
\fmfdot{v2,v3,vx,vxb}
\end{fmfgraph}
\end{center}}
\big)
&=&
U\big(
\parbox{11mm}{\begin{center}
\begin{fmfgraph}(10,4)
\setval
\fmfforce{0w,0h}{v2}
\fmfforce{0.5w,0h}{vx}
\fmfforce{0.5w,6/16h}{vxb}
\fmfforce{0.5w,1h}{vxc}
\fmfforce{1w,0h}{v3}
\fmf{dbl_plain}{v3,v2}
\fmf{boson,right=0.33}{v2,v3}
\fmf{boson}{vx,vxb}
\fmf{dbl_plain,right}{vxb,vxc,vxb}
\fmfdot{v2,v3,vx,vxb}
\end{fmfgraph}
\end{center}}
\big)
+
e C_4\big(
\parbox{7mm}{\begin{center}
\begin{fmfgraph}(6,3)
\setval
\fmfforce{0w,0.5h}{v1}
\fmfforce{1/2w,0.5h}{v2}
\fmfforce{1w,0.5h}{v3}
\fmf{dbl_plain,right}{v2,v3,v2}
\fmf{boson}{v1,v2}
\fmfdot{v1,v2}
\end{fmfgraph}
\end{center}} \big)
U\big(
\parbox{6mm}{\begin{center}\begin{fmfgraph}(5,3)
\setval
\fmfforce{0w,0.5h}{v1}
\fmfforce{0.5w,0.5h}{vx}
\fmfforce{0.5w,0.0h}{vy}
\fmfforce{1w,0.5h}{v2}
\fmf{dbl_plain}{v1,vx,v2}
\fmf{boson,left=1}{v1,v2}
\fmf{boson}{vx,vy}
\fmfdot{v1,v2,vx}
\fmfv{decor.shape=pentagram,decor.filled=1,
      decor.size=2thick}{vy}
\end{fmfgraph}
\end{center}}\big)
\\&&\hspace*{-10mm}
+C_1\big(
\parbox{5mm}{\begin{center}\begin{fmfgraph}(4,3)
\setval
\fmfforce{0w,0.5h}{v1}
\fmfforce{1w,0.5h}{v2}
\fmfforce{0.5w,0.5h}{vx}
\fmfforce{0.5w,0.1h}{vy}
\fmf{dbl_plain}{v1,vx,v2}
\fmf{boson,left=1}{v1,v2}
\fmf{boson}{vx,vy}
\fmfdot{v1,v2,vx}
\end{fmfgraph}
\end{center}}\big)
U\big(
\parbox{7mm}{\begin{center}
\begin{fmfgraph}(6,3)
\setval
\fmfforce{0w,0.5h}{v1}
\fmfforce{1/2w,0.5h}{v2}
\fmfforce{1w,0.5h}{v3}
\fmf{dbl_plain,right}{v2,v3,v2}
\fmf{boson}{v1,v2}
\fmfdot{v1,v2}
\end{fmfgraph}
\end{center}} \big)
+ e C_1\big(
\parbox{5mm}{\begin{center}\begin{fmfgraph}(4,3)
\setval
\fmfforce{0w,0.5h}{v1}
\fmfforce{1w,0.5h}{v2}
\fmfforce{0.5w,0.5h}{vx}
\fmfforce{0.5w,0.1h}{vy}
\fmf{dbl_plain}{v1,vx,v2}
\fmf{boson,left=1}{v1,v2}
\fmf{boson}{vx,vy}
\fmfdot{v1,v2,vx}
\end{fmfgraph}
\end{center}}\big)
C_4\big(
\parbox{7mm}{\begin{center}
\begin{fmfgraph}(6,3)
\setval
\fmfforce{0w,0.5h}{v1}
\fmfforce{1/2w,0.5h}{v2}
\fmfforce{1w,0.5h}{v3}
\fmf{dbl_plain,right}{v2,v3,v2}
\fmf{boson}{v1,v2}
\fmfdot{v1,v2}
\end{fmfgraph}
\end{center}} \big)
U\big(
\parbox{5mm}{\begin{center}
\begin{fmfgraph}(5,2)
\setval
\fmfforce{0w,0.5h}{v1}
\fmfforce{1w,0.5h}{v2}
\fmf{boson}{v1,v2}
\fmfdot{v1}
\fmfv{decor.shape=pentagram,decor.filled=1,
      decor.size=2thick}{v2}
\end{fmfgraph}
\end{center}}
\big).
\end{eqnarray*}

\section{Self-consistent field renormalization}
The case of self-consistent QED is simpler, since
we have the same diagrams as for vacuum QED (with
$S^0$ replaced by $S^\barA$).
Therefore, the
renormalization rules are the same, except that the
free fermion lines become the fermion lines in
the SCF potential $\barA(x)$.

Since there is no tadpole in the fermion propagator
of self-consistent QED,
the only point which is delicate is the
renormalization of tadpoles to calculate
the self-consistent potential.
When all subdivergences are removed, the 
superficial counterterm is obtained by 
expanding the SCF potential in terms
of strong field diagrams and by renormalizing
those. The result is
\begin{eqnarray*}
R^\mu(x;\Gamma) &=&
\barR^\mu(x;\Gamma) + 
e C_4(\Gamma)  \barA^\mu(x).
\end{eqnarray*}
Thus, the counterterm is proportional to
the renormalized potential itself.

\section{Renormalization conditions}
The renormalization theory is an unambiguous method
to remove the subdivergences of a Feynman diagram.
However, once all the subdivergences are removed, we
must still specify the value of the superficial
divergence. This is done by giving renormalization
conditions. In the case of QED without external field,
these conditions are well-known (\cite{Itzykson} p. 413).
However, for QED with external field, they have never
been stated precisely. It is the lack of proper renormalization
conditions that yields the ambiguities in the results
of Dosch and M\"uller \cite{Dosch}.
There are three kinds of renormalization conditions:
for the current, for the photon propagator
and for the electron propagator.
Some of them were investigated in the early days
of QED \cite{Uehling}.

\subsection{The current}
The renormalization condition for the current is deduced
from the neutrality of matter. The total charge of a piece
of matter is the sum of the proton charges and the electron
charges. Thus, the induced charge density due to the
vacuum polarization integrates to zero. If this were not
true, a piece of matter with an equal number of 
electrons and protons would have a net charge.

In the case of strong field QED, we operate the 
renormalization rule for tadpoles (\ref{rentadpole}) with 
${D^0}^{-1}_{\mu\nu}(x,y)$ and
we obtain 
\begin{eqnarray*}
\int\dd y {D^0}^{-1}_{\mu\nu}(x,y) R(\Gamma;y,\nu) &=& 
\int\dd y {D^0}^{-1}_{\mu\nu}(x,y) \barR(\Gamma;y,\nu)
\\&&
 + e C_4(\Gamma)\barj_\mu(x).
\end{eqnarray*}
The left hand side represents the renormalized current
induced by the external current $\barj_\mu(x)$.
Two points must be noticed here. Firstly, the superficial divergence
of the current diagrams (tadpoles) are removed by a
counterterm proportional to the external current,
and any induced current proportional to the external
current cannot be distinguished from a renormalization
of the charge (see \cite{Schweber2} p.553). 
With some hindsight,
this point can be recognized in Ref.\cite{Weisskopf}.
Secondly, the integral of the induced current must be
zero to ensure matter neutrality. Thus
\begin{eqnarray*}
\int\dd x\int\dd y {D^0}^{-1}_{\mu\nu}(x,y) R(\Gamma;y,\nu)=0
\end{eqnarray*}
This second point was clearly made by Uehling \cite{Uehling}.
When the external current is not neutral, this is enough
to determine $C_4(\Gamma)$.

In the case of self-consistent QED, the counterterm
is proportional to $\barA(x)$, and for a neutral atom
the corresponding $\barJ(x)$ integrates to zero. 
Therefore, the criterium of matter neutrality is not
enough to determine the renormalized current of self-consistent
QED. 

\subsection{The photon Green function}
To specify a renormalization condition for
the photon Green function, we 
use the requirement that, very far from
the external field, low frequency 
Compton scattering should be given
by the Thomson formula \cite{Schweber2} 
p.640 and two close charges should 
interact with a Coulomb potential
\cite{Schweber2} p.670 and
\cite{Itzykson} p.325.

In other words, for very large
$\bfx$ and $\bfy$ and small $\bfx-\bfy$,
$\barD_{\mu\nu}(x,y)$ should tend to 
$D^0_{\mu\nu}(x,y)$.

\subsection{The electron Green function}

For the electron Green function, the renormalization
conditions do not seem to have been studied
beyond vacuum QED.
One could probably assume that, 
for very large
$\bfx$ and $\bfy$ and small $\bfx-\bfy$,
$\barS(x,y)$ should tend to
$S^A(x,y)$.
We intend to study the validity of these renormalization 
conditions in the specific case of atomic physics.

\section{Conclusion}
In this paper, the equations of self-consistent 
quantum electrodynamics were derived from
the Schwinger approach. These equations are
valid for any number of electrons in the system
and the functional derivative of the Green
function in the presence of bound electrons was 
used to calculate various perturbative solutions
of the Schwinger equations.

An interesting application of self-consistent
QED is to see if the solitonic solutions
of the Maxwell-Dirac equations \cite{Radford}
survive when the two loop interaction is turned on.
In other words, can an interacting electron be bound
by its own vacuum polarization ?

The general renormalization rules for strong field
QED and self-consistent QED were given. Again,
these rules are valid also in the presence of
bound electrons, because the bound states do not modify
the short-distance or large-momentum asymptotics
of the Green functions.

Much work remains to be done to transform
QED into a practical tool for solid-state
or molecular calculations.
We must investigate more fully the conditions
of renormalization. We should also be able to
calculate the energy-momentum tensor from 
the Green functions, as the current was
written in terms of the electron propagator.
Results in this direction were obtained by
Engel and Dreizler \cite{Engel}, but much
work remains to be done to reach an all-order
renormalization of the QED energy.
The energy is not the only physically interesting 
property and the investigation of the full energy-momentum tensor
might be interesting and 
might provide some physically reasonable
conditions of renormalization.

In this paper, we have only considered
a coherent external source $j_\mu(x)$.
It is also possible to derive Schwinger
equations for a partially coherent source.
The main application of this extension is
the effect of temperature on an electronic
system.

Finally, it would be interesting for
the spectroscopic applications to generalize
the Schwinger equations to a degenerate ``vacuum''.
When the unperturbed state is degenerate, we
must consider not only a single matrix element
$\langle \Phi | A_\mu(x) | \Phi \rangle$
but a matrix
$\langle \Phi_i | A_\mu(x) | \Phi_j \rangle$.
The modifications induced by the presence
of a degenerate vacuum
will be presented in a forthcoming publication.

\section{Acknowledgements}
It is a pleasure to thank Paul Indelicato
and Eric-Olivier Le Bigot for helpful 
discussions. I also thank 
Alessandra Frabetti and
Eric-Olivier Le Bigot for their thorough
reading of the manuscript.
This is IPGP contribution \#0000.

\section{Appendix}
In this appendix, we give the proof of
equations that are used in the text.

\subsection{Derivative of $S^0_N(A)$}

We saw in section \ref{G0N} that $S^0_N(A)= S^0_0(A)+P_N$
with
\begin{eqnarray*}
P_N(x,y) &=& \frac{i}{\hbar c}
\sum_{0<E_n\le E_F} \psi_n(x)\bar\psi_n(y).
\end{eqnarray*}
For notational convenience, we suppress the argument $A$
of the Green functions $S$.
In this equation, the sum is over $N$ states.
The wavefunctions $\psi_n(x)$ are solutions of the Dirac
equation in the presence of $A_\mu(\bfx)$:
\begin{eqnarray*}
\big(i\hbar c \gamma\cdot\partial -mc^2-e\gamma\cdot A(\bfx)\big)\psi_n(x) &=& 0.
\end{eqnarray*}
Therefore 
\begin{eqnarray*}
\big(i\hbar c \gamma\cdot\partial -mc^2-e\gamma\cdot A(\bfx)\big)P_N(x,y) &=& 0,
\end{eqnarray*}
and the adjoint equation, deduced from Eq.(\ref{anothereq})
\begin{eqnarray*}
\big(-i\hbar c \partial^\mu -e A^\mu(\bfx)\big)P_N(y,x)\gamma_\mu
-mc^2 P_N(y,x) &=& 0.
\end{eqnarray*}
We perturb the vector potential into $A_\mu(\bfx)+\epsilon V_\mu(x)$
so that $P_N$ becomes $P_N+\epsilon\delta P_N$, which must be a solution
of the Dirac equation for the perturbed potential. Keeping only the
terms linear in $\epsilon$ we obtain
\begin{eqnarray*}
\big(i\hbar c \gamma\cdot\partial -mc^2-e\gamma\cdot A(\bfx)\big)\delta P_N(x,y) &=& 
e\gamma\cdot V(x) P_N(x,y).
\end{eqnarray*}
The general solution of this equation is
\begin{eqnarray*}
\delta P_N &=& eS^0_0(A)\gamma\cdot V P_N +R_N,
\end{eqnarray*}
where $R_N$ is a solution of the unperturbed Dirac equation.
We repeat the argument for the adjoint equation and we obtain
\begin{eqnarray*}
\delta P_N &=& eS^0_0(A)\gamma\cdot V P_N + e P_N \gamma\cdot V S^0_0(A)
+R'_N,
\end{eqnarray*}
where $R'_N$ is a solution of the Dirac and adjoint Dirac equations.
Finally, we calculate the functional derivative by taking 
$V=\delta_{\lambda,\mu}\delta(x-z)$ so that
\begin{eqnarray*}
\frac{\delta P_N(x,y)}{e\delta A_\lambda(z)} &=& 
S^0_0(x,z)\gamma^\lambda P_N(z,y) 
\\&&
+ P_N(x,z) \gamma^\lambda S^0_0(z,y) +R'_N.
\end{eqnarray*}
To determine $R'_N$ we require that the perturbation does not change
the number of bound electrons. In other words
\begin{eqnarray*}
\Delta N &=&
\int \dd\bfx\, \tr\Big[\gamma^0 
\frac{\delta P_N(x,x)}{\delta A_\lambda(y)}\Big]
=0.
\end{eqnarray*}
To evaluate this, we first calculate
\begin{eqnarray*}
\int \dd\bfx\, \tr[\gamma^0 S^0_0(x,y)\gamma^\lambda P_N(y,x)].
\end{eqnarray*}
We use Eq.(\ref{GN0}) for $E_F=0$, because of the cyclic property
of the trace we can bring the last $\bar\psi_m(x)$ in the front.
then we use the orthogonality of the solutions of the Dirac equation
and we find
\begin{eqnarray*}
\int \dd\bfx\, \tr[\gamma^0 S^0_0(x,y)\gamma^\lambda P_N(y,x)]
&=& \theta(x^0-y^0) 
\\&&\hspace*{-30mm}\times
\sum_{0<E_m\le E_F} \tr[\bar\psi_m(y)\gamma^\lambda \psi_m(y)].
\end{eqnarray*}
The same equation for the second part ($P_NS^0_0$) gives the 
same expression,
where $\theta(x^0-y^0)$ is replaced by $\theta(y^0-x^0)$. Thus
\begin{eqnarray*}
\int \dd\bfx\, \tr[\gamma^0 (S^0_0(x,y)\gamma^\lambda P_N(y,x)
                  +P_N(x,y)\gamma^\lambda S^0_0(y,x))]
&&\\\hspace*{-60mm}
= \sum_{0<E_m\le E_F} \tr[\bar\psi_m(y)\gamma^\lambda \psi_m(y)].
\end{eqnarray*}
Now if we take
$R'_N(x,y;z,\lambda)=P_N(x,z)\gamma^\lambda P_N(z,y)$ we obtain
\begin{eqnarray*}
\int \dd\bfx \tr[\gamma^0 R'_N(x,x;y,\lambda)] 
&=& -\sum_{0<E_m\le E_F} \tr[\bar\psi_m(y)\gamma^\lambda \psi_m(y)].
\end{eqnarray*}
which compensates exactly for the previous term.
Finally, since we know that
\begin{eqnarray*}
\frac{\delta S^0_0(x,y;A)}{\delta A_\lambda(z)} &=&
S^0_0(x,z;A)\gamma^\lambda S^0_0(z,y;A).
\end{eqnarray*}
we have obtained that
\begin{eqnarray*}
\frac{\delta S^0_N(x,y;A)}{\delta A_\lambda(z)} &=&
S^0_N(x,z;A)\gamma^\lambda S^0_N(z,y;A).
\end{eqnarray*}
This is a very satisfactory result, which shows that the 
perturbative solution of the Schwinger equation has the same
form for the vacuum and the charged cases.

\subsection{Integral}
In this section we show the following identity
\begin{eqnarray}
F(\varphi) &=& F(0) + \int_0^1 \dd\lambda \int \dd x 
\frac{\delta F(\lambda\varphi)}{\delta\varphi(x)}\varphi(x).
\label{appendiceIntegral}
\end{eqnarray}
This is a functional form of the classical Taylor
formula
\begin{eqnarray}
f(x) &=& f(0) + \int_0^1 \dd\lambda f'(\lambda t) t, \label{Taylor1}
\end{eqnarray}
where $f'(t)$ is the derivative of $f(t)$.
The $n$-dimensional generalization of Eq.(\ref{Taylor1}) is
\begin{eqnarray}
f(\bfx) &=& f(0)+\int_0^1 \dd\lambda \frac{\dd}{\dd\lambda}
   f(\lambda\bfx)\nonumber\\
&=& f(0) + \int_0^1 \dd\lambda \sum_{i=1}^n
 \partial_i f(\lambda \bfx) x^i.
\label{ndim}
\end{eqnarray}

Eq.(\ref{appendiceIntegral}) can be derived as the
above $n$-dimensional case, but wee choose to prove it by an explicit
calculation which shows that the purpose of the integral 
over $\lambda$ is to change the multiplicity of some terms.
We start from the Taylor 
expansion for functional derivatives
\begin{eqnarray*}
F(\varphi) &=& F(0) + \sum_{n=1}^\infty \frac{1}{n!}
\int \dd y_1 \cdots\dd y_n 
\frac{\delta^n F(0)}{\delta\varphi(y_n)\cdots \delta\varphi(y_1)}
\\&&\times
\varphi(y_1)\cdots \varphi(y_n).
\end{eqnarray*}
The notation used in this equation means that the functional
derivatives of $F$ are taken at $\phi=0$.
Therefore, using the symmetry of functional derivatives
with respect to their arguments
\begin{eqnarray*}
\frac{\delta F(\varphi)}{\delta\varphi(x)} &=& 
\sum_{n=1}^\infty \frac{1}{(n-1)!}
\int \dd y_1 \cdots\dd y_{n-1} 
\\&&\times
\frac{\delta^n F(0)}{\delta\varphi(x)\delta\varphi(y_{n-1})
\cdots \delta\varphi(y_1)}
\varphi(y_1)\cdots \varphi(y_{n-1}).
\end{eqnarray*}

In Eq.(\ref{appendiceIntegral})
the purpose of the integral over $\lambda$ 
is to replace the
factor $1/(n-1)!$ by the correct factor $1/n!$.
\begin{eqnarray*}
\frac{\delta F(\lambda\varphi)}{\delta\varphi(x)} &=&
\sum_{n=1}^\infty \frac{\lambda^{n-1}}{(n-1)!}
\int \dd y_1 \cdots\dd y_{n-1}
\\&&\times
\frac{\delta^n F(0)}{\delta\varphi(x)\delta\varphi(y_{n-1})
\cdots \delta\varphi(y_1)}
\varphi(y_1)\cdots \varphi(y_{n-1}).
\end{eqnarray*}
The notation $\delta F(\lambda\varphi)/\delta\varphi(x)$
has the following meaning.
The function $G(x,\varphi)=\delta F(\varphi)/\delta\varphi(x)$
is a function of two independent variables ($x$ and $\varphi$).
Then 
$\delta F(\lambda\varphi)/\delta\varphi(x)=G(x,\lambda\varphi)$.

Therefore,
\begin{eqnarray*}
\int_0^1 \dd\lambda \int \dd x 
\frac{\delta F(\lambda\varphi)}{\delta\varphi(x)}\varphi(x)
&=&
\sum_{n=1}^\infty \frac{1}{n!}
\int \dd y_1 \cdots\dd y_{n-1}\dd x
\\&&\times
\frac{\delta^n F(0)}{\delta\varphi(x)\delta\varphi(y_{n-1})
\cdots \delta\varphi(y_1)}
\\&&\times
\varphi(y_1)\cdots \varphi(y_{n-1})\varphi(x)
\\&=& F(\varphi)-F(0).
\end{eqnarray*}

Note that an obvious consequence of 
Eq.(\ref{appendiceIntegral}) is that 
\begin{eqnarray}
\frac{\delta}{\delta\varphi(x)} \int_0^1 \dd\lambda \int \dd x 
\frac{\delta F(\lambda\varphi)}{\delta\varphi(x)}\varphi(x)
&=& \frac{\delta F(\varphi)}{\delta\varphi(x)}.
\label{appendicederive}
\end{eqnarray}

\end{fmffile}

\end{document}